 \documentclass[final,5p,times,twocolumn]{elsarticle}

\usepackage{amssymb}

\usepackage{amsmath}

\usepackage{lipsum}

\usepackage{enumitem}

\usepackage{graphicx}
\usepackage{caption}
\usepackage{eurosym}

\usepackage{xurl}
\usepackage{hyperref}

\usepackage{booktabs}

\usepackage{pdflscape}

\journal{Applied Energy}

\begin{document}

\begin{frontmatter}


\title{Direct and Indirect Hydrogen Storage: Dynamics and Interactions in the Transition to a Renewable Energy Based System for Europe}


\author[first]{Zhiyuan Xie\corref{cor1}}
\ead{zyx@mpe.au.dk}
\author[first,second]{Gorm Bruun Andresen}

\cortext[cor1]{Corresponding author.}

\affiliation[first]{organization={Department of Mechanical and Production Engineering, Aarhus University},
            city={Aarhus},
            country={Denmark}}
            
\affiliation[second]{organization={iCLIMATE Interdisciplinary Centre for Climate Change, Aarhus University},
            city={Aarhus},
            country={Denmark}}


\begin{abstract}

To move towards a low-carbon society by 2050, understanding the intricate dynamics of energy systems is critical. Our study examines these interactions through the lens of hydrogen storage, dividing it into ‘direct’ and ‘indirect’ hydrogen storage. Direct hydrogen storage involves electrolysis-produced hydrogen being stored before use, while indirect storage first transforms hydrogen into gas via the Sabatier process for later energy distribution. Firstly, We utilize the PyPSA-Eur-Sec-30-path model to capture the interactions within the energy system. The model is an hour-level, one node per country system that encompasses a range of energy transformation technologies, outlining a pathway for Europe to reduce carbon emissions by 95\% by 2050 compared to 1990, with updates every 5 years. Subsequently, we employ both quantitative and qualitative approaches to thoroughly analysis these complex relationships. Our research indicates that during the European green transition, cross-country flow of electricity will play an important role in Europe's rapid decarbonization stage before the large-scale introduction of energy storage. Under the paper cost assumptions, fuel cells are not considered a viable option. This research further identify the significant impact of natural resource variability on the local energy mix, highlighting indirect hydrogen storage as a common solution due to the better economic performance and actively fluctuation pattern. Specifically, indirect hydrogen storage will contribute at least 60\% of hydrogen storage benefits, reaching 100\% in Italy. Moreover, its fluctuation pattern will change with the local energy structure, which is distinct difference with the unchanged pattern of direct hydrogen storage and battery storage.

\end{abstract}




\end{frontmatter}




\section{Introduction}
\label{introduction}

The global energy landscape is undergoing significant challenges due to escalating demands from a booming population \cite{mutschler2021benchmarking}. As evidence of climate change's adverse impacts accumulates, the urgency of transitioning to greener energy sources becomes urgent \cite{aghahosseini2023energy}. The discourse on this green transition often emphasizes the integration of energy systems, which can enhance energy efficiency and mitigate decarbonization expenses \cite{gea2021role, li2022cost}. For instance, the incorporation of wind-solar hybrid systems enhances both diurnal and seasonal energy accessibility, minimizing the need for extensive energy storage investments \cite{renspatial2019, gangopadhyay2024wind}. Similarly, combined heat and power systems amplify the overall efficiency of the energy system \cite{erixnoenergy2022, wang2015modelling}. Nevertheless, the integration of diverse renewable technologies introduces complex interactions within the energy system. In this paper, we seek to improve the comprehension of intricate dynamics in these new complex energy systems, specifically through the lens of hydrogen in its direct and indirect storage forms. Our goal is to use this understanding to guide and facilitate the green transition in a broad sense.

At this juncture, it’s essential to review current literature to grasp how different studies perceive the interplay within the energy system, especially among renewable energy, storage, and electricity transmission. The electricity sector is witnessing transformative changes with the integration of renewable energy sources (RES), especially wind and solar \cite{alshetwisustainable2022, liangemerging2017}. Historically, fossil fuel costs were the primary determinants of electricity prices. However, as RES adoption grows, this paradigm is shifting \cite{borensteinprivate2012, joskowcomparing2011}. One of the notable challenges introduced by these renewables is their intermittency, which impacts electricity pricing \cite{gonzalezdesigning2023, fang2021characteristics}. In situations where wind and solar outputs peak, the marginal cost of electricity generation drastically decreases. This has led to moments of ultra-low, and sometimes even negative, electricity prices \cite{bibernegative2022, devosnegative2015}. De Vos \cite{devosnegative2015} suggests that the shift towards renewables reduces dependence on fossil fuels, potentially lowering electricity prices, especially in competitive market regions. Implementing CO$_2$ allowance prices emerges as a solution to counter negative market prices. Furthermore, the inability to store large amounts of electricity aggravates pricing challenges. Such limitations induce price volatility, underscoring the intricate dance of balancing electricity supply with demand \cite{shaofeature2021}.

To facilitate and advance the large-scale integration of renewable energy into the energy system, energy storage is increasingly seen as an indispensable solution, emerging as a central pillar of the contemporary energy framework \cite{solomon2014role, mcpherson2018role}. Currently, pumped hydropower storage dominates the global energy storage landscape. However, its potential is somewhat restricted due to terrain prerequisites, a limitation particularly evident in Europe where most of its potential has been tapped \cite{gimenogutierrezassessment2015}. Batteries often serve as the preferred choice for short-duration energy storage, whereas hydrogen (H$_2$) is seen as a promising long-duration storage alternative \cite{scholzapplication2017, cebullaelectrical2017}. Areas that face pronounced electricity price fluctuations, owing to the variability of renewable energy sources, typically witness a spike in investments geared towards storage infrastructure \cite{schmidtprojecting2019}. However, the role of energy storage in green transitions isn't universal; its significance can differ based on various sector coupling scenarios, as meticulously explored by Victoria \textit{et al} \cite{victoriarole2019}. This evolving landscape underscores the intertwined nature of renewable energy generation, electricity pricing, and energy storage.

Cross-country flow of electricity is a pivotal factor differentiating various energy scenarios. While many existing research models overlook the intricacies of interconnections between countries or regions \cite{gotskecost2023}, the benefits of such interconnections are undeniable. Enhancing cross-country electricity flow and sector coupling simultaneously can create synergies that yield significant benefits \cite{brownsynergies2018}. Enhanced power transmission not only alleviates grid congestion but also ensures the smooth integration of renewable energy, subsequently influencing electricity prices. Moreover, robust interconnections pave the way for more streamlined electricity markets, fostering competitive pricing. Such pricing structures not only benefit consumers but also guarantee consistent electricity supplies \cite{iranmehrmodeling2022, shenimpacts2022, schlachtbergerbenefits2017}. Even in scenarios where bolstering transmission is viewed with skepticism, either due to perceived challenges or economic concerns \cite{caoprevent2021}, the significance of current transmission capabilities and the potential for expansion should not be sidelined.

Current research extensively explores the interplay between different components of the energy system, while often under lots of restrictive assumptions. This focus predominantly on local connections between a few energy sectors results in a narrow view, largely overlooking the broader, interconnected dynamics of the energy landscape \cite{gotskecost2023}. Such a limited perspective tends to underestimate the potential of some renewable energy technology by not fully accounting for its capabilities and impact when integrated on a larger scale. Consequently, there's a pressing need for studies that transcend these local confines to offer a more holistic understanding of the energy system. To address these limitations and provide a more encompassing view of the energy transition, this paper will utilize the PyPSA-Eur-Sec-30-path model developed by Victoria \textit{et al} \cite{victoriaearly2020}. This model offers valuable insights into the overall energy transition process by presenting a scenario of transitioning to high renewable energy reliance within the context of interconnectedness across 30 European countries. It encompasses a range of necessary renewable energy-related technologies and delves into the impacts of different CO$_2$ emission reduction trajectories on green transition costs. Their research concludes that initiating emission reductions early and maintaining a steady pace is more cost-effective than delayed, rapid reductions. Building on this finding, this paper focuses on the nuanced interactions observed under the strategy of early and consistent emission reduction.

Based on having a comprehensive vision for the transformation of the energy system, this paper utilizes hydrogen storage as a key entry point to unravel the complex interactions within the system, due to its significant intersections with the heating, power, and gas sectors. In line with this vision, this paper introduces innovative concepts of ‘direct’ and ‘indirect’ H$_2$ storage, pivotal in understanding H$_2$'s integral function across these diverse sectors. This conceptual framework enables the study to expand further, exploring a range of technologies within the energy model and highlighting the multifaceted nature of energy transitions. To analyze the dynamics of various renewable technologies comprehensively, this research employs a combination of Fast Fourier Transform (FFT), Continuous Wavelet Transform (CWT), and Cycle Capture (CC) methods. These approach deepens our understanding of the time-dependent behavior and interactions of different elements in the energy system.

The structure of this paper is as follows: Section~\ref{methods} provides a concise introduction to the research model and methods employed and elucidates the concepts of direct and indirect hydrogen storage. A detailed mathematical description of the model is available in the ~\ref{appendix a} for further reference. Section~\ref{results-and-discussion} ventures into an exploration of the intricate interdependencies within the energy system, offering insights from various analytical angles. While section 3 includes select illustrative figures, a full set of corresponding diagrams is housed in ~\ref{appendix b} for completeness. Finally, Section~\ref{conclusion} summarizes the key findings and conclusions drawn from this study.

\section{Methods}
\label{methods}

\subsection{Model Description and Hydrogen Storage}
\label{subsec:model-description}

The PyPSA-Eur-Sec-30-path model \cite{victoriaearly2020} stands as an open-source network model of the European energy system, featuring hourly resolution and a single-node representation for each country. In this paper, the scope of the model includes the electricity, heating, hydrogen and gas sectors. The full network consists of 30 nodes, representing the 28 EU member states as of 2018, excluding the islands of Malta and Cyprus. However, it extends to include the well-connected non-EU countries Norway, Switzerland, Serbia, and Bosnia-Herzegovina. Each of these neighboring countries is interconnected through high voltage transmission lines. 

The PyPSA-Eur-Sec-30-path model employ a brownfield optimization approach, which includes existing power plant capacities until the end of their technical lifetime. In addition to these, the model may invest in new energy infrastructure to cover the energy demands. To analyze a decarbonization pathway for the European power system, a myopic approach is used. This approach, characterized by limited foresight over the investment period \cite{ponceletmyopic2016}. This approach may incur higher overall system costs compared to perfect foresight optimization due to stranded investments.

\begin{figure*}[t]
	\centering 
	\includegraphics[width=0.85\textwidth, angle=0]{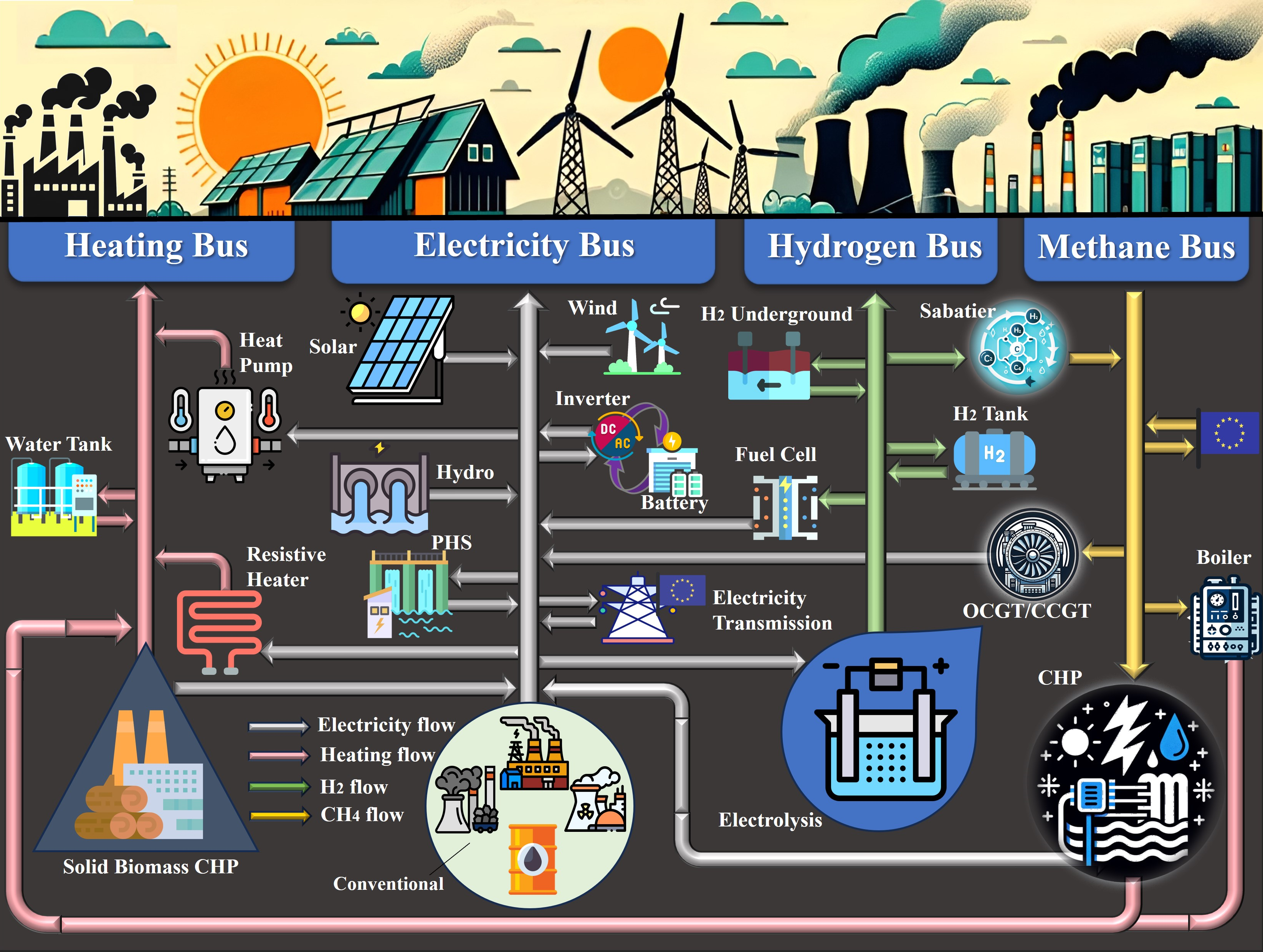}	
	\caption{Energy Flow Dynamics in Country-Specific Nodes: An Analysis of Electricity, Heating, Hydrogen, and Methane Buses with Technological Correlations. (Icons sourced from Flaticon.com and OpenAI, available for free use).} 
	\label{fig_1}
\end{figure*}

Figure~\ref{fig_1} represent a schematic of a multi-vector energy system, integrating various sources and forms of energy into a cohesive network. That is, the schematic diagram of PyPSA-Eur-Sec-30-path model. The system is organized into four primary ‘buses’, each corresponding to distinct energy carriers: Heating, Electricity, Hydrogen, and Methane, illustrating various technologies and processes within.

Central to the system, the Electricity Bus interfaces with renewable sources, including solar, wind, and hydroelectric stations, alongside solid biomass. It also encompasses conventional power plants powered by oil, lignite, coal, or nuclear energy, highlighting the synergy between traditional and renewable energy sources. The inclusion of storage solutions—such as a combination of inverters and batteries, pumped-hydro storage (PHS), and hydrogen storage - underscores their importance in a renewable-dominant energy landscape. The Electricity Bus serves as a nexus, supplying or receiving energy from other buses and nations. It interconnects with the Methane Bus via Sabatier technology, cycling methane back through various power generation technologies. Additionally, it distributes electricity to heating applications and connects with other national grids through high-voltage direct current (HVDC) systems. Notably, the conventional power plants, hydroelectric stations, and PHS are exogenously fixed in the model. All except hydroelectric stations, and PHS are subject to decommissioning upon reaching their technical lifespan.

The Heating Bus includes two different configurations, depending on population density: one for urban heating and another for rural heating. In urban areas with high population density, a variety of technologies are integrated into district heating networks to supply heating. These include central ground-sourced heat pumps, heat resistors, gas boilers, and combined heat and power (CHP) units. Additionally, individual air-sourced heat pumps are available as an option in these urban settings. Conversely, in rural, low-density population areas, heating is typically provided by individual installations, such as ground-sourced heat pumps, heat resistors, and gas boilers, catering to the specific needs of these less populated regions.

The Hydrogen Bus is characterized by electrolysis units that produces green hydrogen using electricity, presumably from renewable sources in the late Europe green transition stage. While underground storage represents a large-scale solution, geological constraints necessitate the inclusion of above-ground H$_2$ storage tanks within the model. The hydrogen flow is dynamic, serving either storage or immediate conversion to methane via the Sabatier process. Hydrogen may also be provided to fuel cells for direct conversion to electricity.

Lastly, the Methane Bus features the Sabatier process for synthesizing natural gas from hydrogen and carbon dioxide, with CHP units indicating methane's role in both heating and power generation. Boilers for heating and turbines for electricity production further highlight the dual utility of methane in this integrated system. This illustrates that methane, derived from the Hydrogen Bus, serves a dual purpose: it meets demands for both heating and electricity. And we can conclude hydrogen may operate in our model through 5 distinct pathways, which are:

\begin{enumerate}[label=\arabic*),ref=\arabic*)]
  \item \textbf{Electrolysis - Store - Fuel Cell (ESFC):} Hydrogen is produced, stored, and then used in fuel cells to generate electricity that meets power demands on the electricity bus.
  \item \textbf{Electrolysis - Store - Sabatier - Electricity (ESSE):} Hydrogen is produced, stored, and then used in the Sabatier reaction to create gas that powers Combined Heat and Power (CHP) units, Open-Cycle Gas Turbines (OCGT), or Combined-Cycle Gas Turbines (CCGT), generating electricity to fulfill electricity bus requirements.
  \item \textbf{Electrolysis - Store - Sabatier - Heating (ESSH):} Hydrogen is produced, stored, and then used in the Sabatier reaction to produce gas for CHP systems or boilers, providing heat to satisfy heating demands.
  \item \textbf{Electrolysis - Sabatier - Store - Electricity (ESE):} Hydrogen produced through electrolysis is utilized in the Sabatier reaction to synthesize gas. This gas is then stored and subsequently dispatched as needed for electricity generation through CHP, OCGT, and CCGT, which in turn supply the electricity bus.
  \item \textbf{Electrolysis - Sabatier - Store - Heating (ESH):} Following electrolysis, hydrogen is immediately processed via the Sabatier reaction to produce gas. This gas can either be stored directly in gas storage or immediately used in CHP and boilers. The energy generated by these systems can then be stored as heat in thermal storage units. Finally, the heating energy will provide to the heating bus.
\end{enumerate}

And here is no doubt that the pathway 1 to 3 belongs to the H$_2$ storage, and we called it ‘direct’ H$_2$ storage. However, the pathway 4 and 5 both involves the electrolysis, which can also be considered a form of storage since it utilizes surplus electricity from the electricity bus to store energy in another form \cite{sgobbihow2016}. Therefore, we categorize the pathway 4 and 5 as ‘indirect’ H$_2$ storage, highlighting their role in energy conversion and storage.

Overall, the schematic illustrates a complex, integrated energy system that leverages multiple energy carriers and storage solutions to maximize efficiency and utilize renewable energy sources. For an in-depth exploration of the model and technologies involved, such as the COP of heat pump, land use constraints and the detail of the solid biomass, the reader is directed to Victoria \textit{et al}. \cite{victoriaearly2020}.

\subsection{Basis for Country Case Selection}
\label{subsec:basis-for-country-case-selection}

\begin{figure*}[t]
	\centering 
	\includegraphics[width=0.9\textwidth, angle=0]{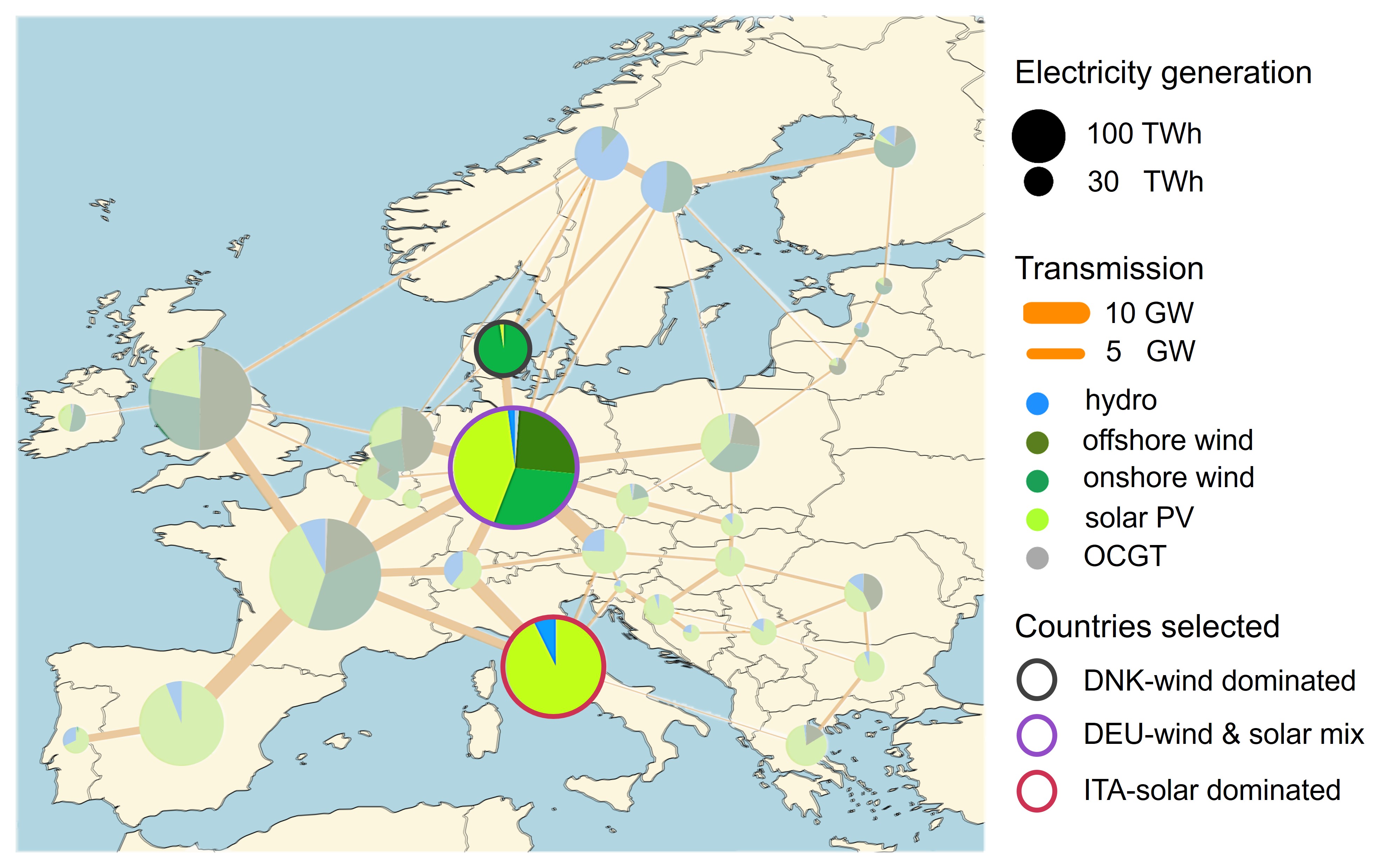}	
	\caption{Diverse Energy Mixes in European Electricity Generation by 2050 (Denmark, Germany, Italy Highlighted), adapted from Victoria \textit{et al}. \cite{victoriaearly2020}, Figure 6.} 
	\label{fig_2}
\end{figure*}

Diverse geographical landscapes and natural resource endowments across Europe lead to varying renewable energy potentials among different countries. As Figure~\ref{fig_2} illustrates, by 2050, Denmark (DNK) is expected to harness predominantly wind energy, Germany (DEU) to utilize a balanced mix of wind and solar, and Italy (ITA) to rely heavily on solar power. This projection underscores the distinct energy generation methods that each country will likely adopt. Such differences are crucial for this paper, it means the energy transition path in Denmark, Germany, and Italy emerges as typical or extreme subjects to represent the spectrum of energy transition outcomes from 2020 to 2050. Moreover, A comparative analysis of these distinct countries will specifically account for the influence of natural resource variability on the energy transition, thereby enriching our understanding with a more holistic perspective. 

Due to space constraints, the Results and Discussion section of this paper will primarily focus on graphical relevant to Germany. Comprehensive graphical for Denmark and Italy, while not featured in the main section, are included in the ~\ref{appendix b} for thorough examination. The analysis in this section will commence with the findings pertinent to Germany, followed by detailed examinations of Denmark and Italy, ensuring a comprehensive discussion across all three countries.

\subsection{Mathmatical Modelling}
\label{subsec:mathmatical-modelling}

For a detailed explanation of the mathematical foundation of the PyPSA-Eur-Sec-30-path model, please refer to ~\ref{appendix a}. This appendix details the model's minimum cost objective function, the equilibrium constraints for supply and demand, and the pathway for CO$_2$ emissions reduction.

It is essential to emphasize that the electricity prices at specific nodes (with one node representing each country) are derived from the power supply and demand balance constraints at those respective nodes, meaning the electricity price used in this study are locational marginal prices (LMP). The same pricing principle applies to hydrogen. Similarly for the hydrogen price. From an economic perspective, this marginal price indicates the additional cost incurred for delivering one more unit of electricity or hydrogen to a particular node. It's crucial to recognize that this price is not a standalone number, it is the result of the dynamic interactions between multiple elements of the energy system. 

In the Mathematical Modelling section, the primary focus will be on the mathematical formulation of the economic indicators and the qualitative analysis methods employed.

\subsubsection{Mathematical Formulations for Economic Metrics}
\label{subsubsec:mathematical-formulations-for-energy-system-performance-metrics}

For the economic indicators, we mainly examine the Levelized Cost of Storage (LCOS), Unit Benefit (UB), overall price spread (OPS) and Cycle Frequency (CF). The Levelized Cost of Storage (LCOS) mathematical expressions are as follows \cite{parzenbeyond2022}:

\begin{equation}
    LCOS = \frac{\sum_{t=0}^{T} C_{\text{total cost}, t}}{\sum_{t=0}^{T} E_{\text{dis}, t}}
\end{equation}

\noindent The term $\sum_{t=0}^{T} C_{\text{total cost}, t}$ represents the accumulated annualized cost, which includes consideration of the discount rate at $7\%$. For storage, the total cost typically encompasses the accumulated annualized cost in parts of the charging link (such as electrolyzers or battery inverters), the storage system itself, and the discharging link (such as fuel cells). However, for H$_2$ work pathways 4 and 5, since the H$_2$ is transported directly from electrolysis to the Sabatier process, the accumulated annualized cost for the storage itself is not included in the calculation of the total cost. It is important to note that $\sum_{t=0}^{T} E_{\text{dis}, t}$ refers to the amount of energy available for discharge, calculated prior to considering discharge efficiency. This approach facilitates a straightforward comparison across different H$_2$ work pathways and battery systems by placing them on a uniform benchmark.

The Unit Benefits (UB) mathematical expressions are as follows:

\begin{equation}
    UB = \frac{\sum_{t=0}^{T} R_{\text{cost}, t} - \sum_{t=0}^{T} C_{\text{charging cost}, t}}{\sum_{t=0}^{T} E_{\text{dis}, t}}
\end{equation}

\noindent In this context, $\sum_{t=0}^{T} R_{\text{cost}, t}$ refers to the revenue generated from the sale of H$_2$ to the Sabatier process or from selling electricity to the electricity grid. $\sum_{t=0}^{T} C_{\text{charging cost}}$, on the other hand, denotes the cost incurred from purchasing electricity from the electricity bus.

The overall price spread (OPS) mathematical expressions are as follows:

\begin{equation}
\label{equation_3}
    OPS = \frac{1}{N} \sum_{c=1}^{N} (P_{\text{sell}, c} - P_{\text{buy}, c})
\end{equation}

\noindent While $P_{\text{sell}, c}$ is the selling price per unit of energy during cycle c, $P_{\text{buy}, c}$ is the selling price per unit of energy during cycle $c$, $N$ is the total number of cycles detected by the Cycle Capture (CC) method over a year. For detailed information on the CC method and its mechanism for collecting cycle data, please see Section~\ref{subsec:integration-of-FFT, CWT, and-CC-methods-for-advanced-analysis}.

\subsubsection{Mathematical Formulations for Qualitative Analysis Methods}
\label{subsubsec:mathematical-formulations-for-qualitative-analysis-methods}

During this stage, we apply a combination of Fast Fourier Transform (FFT), Continuous Wavelet Transform (CWT), and Cycle Capture (CC) methodologies. These techniques are employed to meticulously analyze time series data pertaining to storage capacities, renewable energy generation, electricity pricing, and the marginal price of hydrogen. However, the detailed exposition of the Cycle Capture (CC) method is reserved for Section~\ref{subsec:integration-of-FFT, CWT, and-CC-methods-for-advanced-analysis}.

The Fast Fourier Transform (FFT) and the Continuous Wavelet Transform (CWT) are key mathematical tools for analyzing and interpreting data volatility. FFT decomposes a time-based signal into its constituent frequencies, facilitating the detection of frequency components within time series data. On the other hand, CWT provides a more detailed view of non-stationary signals, with frequency components that change over time \cite{zhuresearch2011}. However, it's important to note that CWT may emphasize larger scales (lower frequencies) more, as the wavelets at larger scales cover a broader range in frequency space, thus potentially dominating the energy of the CWT, especially when using a logarithmic scale spacing \cite{rodenaswavelet1997}. While FFT can correctly determine the main frequency of the data, it has resolution limitations in time-frequency representation, and assumes that the frequency components of the signal do not vary over time \cite{zhuresearch2011}. Consequently, it cannot provide the exact time and frequency of the signal simultaneously. In this study, we employ FFT to capture the frequency characteristics of time series data, and CWT to reveal the internal frequency variation over time.

The Fast Fourier Transform (FFT) mathematical expressions are as follows:

\begin{equation}
    X(k) = \sum_{m=0}^{J-1} x(m) \cdot e^{- \frac{2 \pi i}{J} km}
\end{equation}

\noindent Where $X(k)$ is the output of FFT for each value of $k$, $x(m)$ means the input time series data, $J$ represents the number of samples.

The Continuous Wavelet Transform (CWT) mathematical expressions are as follows \cite{stepanovconstruction2017}:

\begin{equation}
    X_{w}(a, b) = \frac{1}{|a|^{1/2}} \int_{-\infty}^{\infty} x(t) \psi \left( \frac{t-b}{a} \right) dt
\end{equation}

\noindent where $\psi \left( t \right)$ is a continuous function known as the mother wavelet, and the overline represents the complex conjugate operation. And the mother wavelet used in this paper is the Morlet wavelet. The Morlet wavelet is advantageous for its balance between time and frequency localization, offering a good compromise for analyzing signals with time-varying frequencies \cite{Guo2022Wavelet}. And the mathematical expressions are as follows:

\begin{equation}
\psi(t) = e^{-t^2/2} \cos(5t)
\end{equation}

\subsection{Integration of FFT, CWT, and CC Methods for Advanced Analysis}
\label{subsec:integration-of-FFT, CWT, and-CC-methods-for-advanced-analysis}

The combination of CWT and FFT offers numerous benefits, enabling the identification of dominant frequencies within the data and tracking how these frequencies vary over time. However, when the large-scale features in the data are prominent, smaller scale features, such as the diurnal fluctuations (24h) of H$_2$, may be diminished, while that of Battery would not be affected. 

To address this, we introduce a method similar to rainflow-counting method to capture cycles in the time series data, termed the Cycle Capture (CC) method, complementing FFT and CWT. In the CC method developed by this paper, users have the flexibility to adjust data filtering thresholds, as well as rising (charging) and falling (discharging) thresholds, to effectively capture cycle information. For instance, in our study on tracking cycle information in the variation of the battery storage filling level data over a year, only the processes where charging and discharging exceed $10\%$ of the normalized value are identified together as a cycle. Notably, this method is capable of handling data that initiates in any state—waiting, charging, or discharging—ensuring that no initial state or relevant information is overlooked in cycle detection. For a detailed understanding of the CC method's operational logic, please refer to Figure~\ref{fig_S1}. Additionally, the source code for the CC method can be found in Section~\ref{subsec:code-availability}.

CC provides two primary advantages in synergy with FFT and CWT:

\begin{enumerate}[label=\arabic*),ref=\arabic*)]
  \item Short-Term Fluctuation Detection: The CC method is adept at capturing minor fluctuations in data, mainly observed on daily to weekly time scales, due to the small threshold used in this study. This ability ensures that while FFT and CWT identify dominant long-term or large-scale patterns in data, CC pinpoints the smaller, more nuanced fluctuations, providing a comprehensive view of both macro and micro data trends.
  \item Result Verification: When dominant patterns in the data predominantly consist of short cycles, results from FFT, CWT, and CC might converge, highlighting these frequent short-term fluctuations. Such convergence signifies the robustness of the findings, as short-term results from FFT and CWT can be cross verified with CC. This cross validation not only emphasizes commonalities but also affirms the selected threshold's appropriateness for CC, ensuring that all methods align in their analytical objectives.
\end{enumerate}

In short, the synergistic application of FFT, CWT, and CC offers a comprehensive understanding of the data's true volatility, aiding in the exploration of interactions within various sectors of the energy system. Specifically, FFT identifies the primary frequency of the data, CWT traces frequency variations over time, and CC emphasizes short-term frequency information potentially overlooked by FFT and CWT. Moreover, when the dominant frequency exhibits short-term fluctuations, both FFT and CWT can validate the appropriateness of the parameter settings in CC.

Besides the benefits gained from integrating FFT and CWT, CC method is independently used in this study to investigate the intricate interactions within the energy system. This is because CC effectively tracks the start and end index of each cycle, laying the groundwork for the calculations in Equation ~\ref{equation_3}.

\subsection{Key Parameter}
\label{subsec:key parameter}

\subsubsection{Input and Output Data}
\label{subsubsec:input-and-output-data}

In this study, the load side demand is determined using a variety of 2015 data sources. Hourly electricity demand for each country is derived from data provided by the EU Network Transmission System Operators of Electricity (ENTSO-E), which is further refined by the Open Power System Data (OPSD) initiative. Additionally, annual heat demand is converted to an hourly resolution using population-weighted temperature time series. For the generation side, country-specific onshore and offshore wind capacity factors are modeled by transforming wind velocity data from the Climate Forecast System Reanalysis (CFSR) into wind generation estimates. Similarly, for photovoltaic (PV) generation, it is assumed that $50\%$ of the PV capacity consists of rooftop installations. Land use considerations involve aggregating suitable land in each reanalysis grid cell based on the Corine Land Cover (CLC) database, while excluding Natura 2000 protected areas. This approach draws on the comprehensive data source overviews provided in studies by Brown \textit{et al}., Zhu \textit{et al}. \cite{victoriaearly2020, brownsynergies2018, zhuimpact2019}. Winterscheid \textit{et al}. \cite{winterscheidintegration2017}, who emphasize the importance of data quality for result accuracy. The input data used in our model are sourced from open, transparent, and extensively verified datasets. Both the input datasets and the model-generated data are publicly available at the repository: \url{https://zenodo.org/records/4010644}.

\begin{figure}
	\centering 
	\includegraphics[width=0.45\textwidth, angle=0]{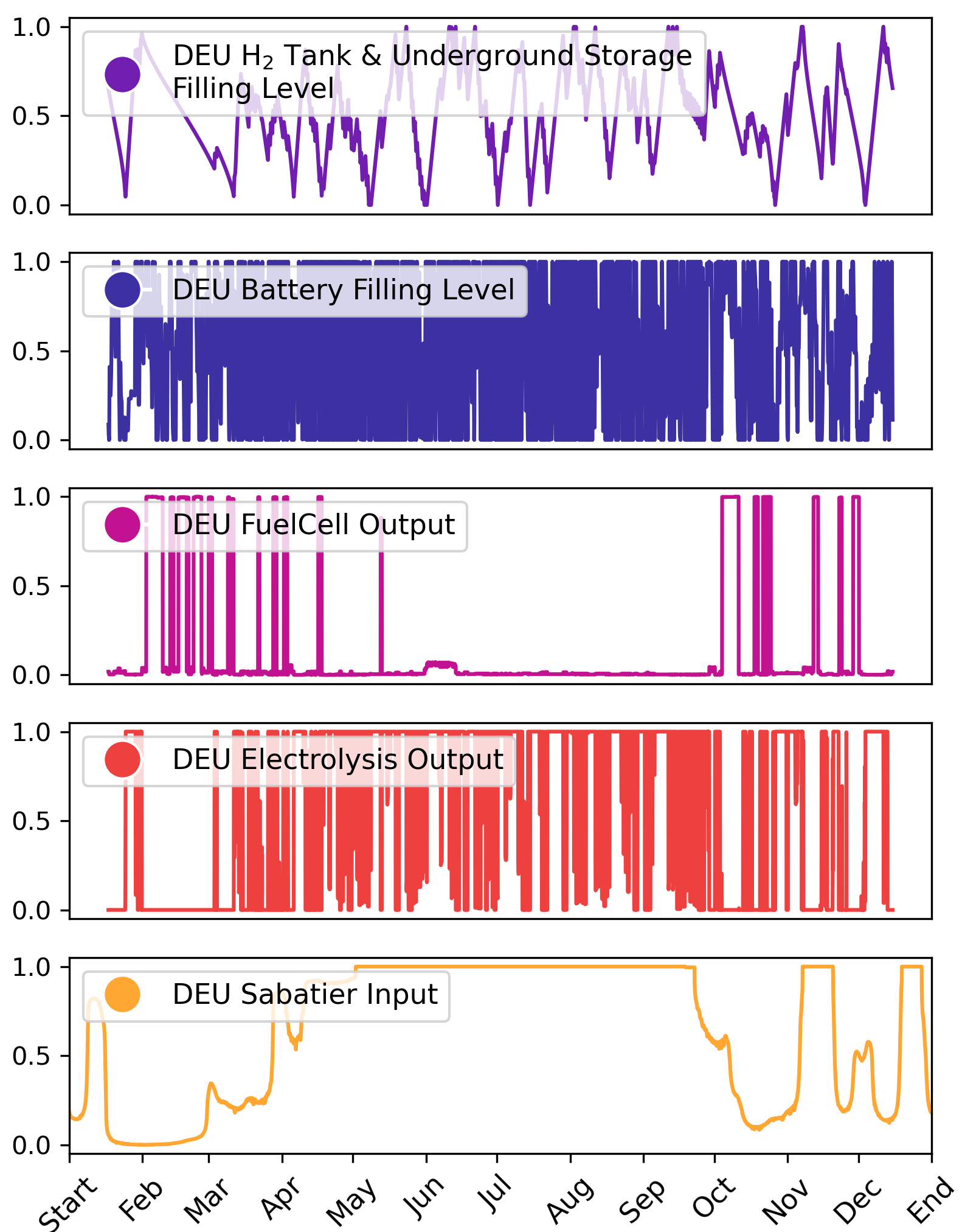}	
	\caption{Time Series Analysis of Germany's Hydrogen Sector and Battery Storage for 2050} 
	\label{fig_3}
\end{figure}

Figure~\ref{fig_3} offers a detailed visualization of the technological original time series data within Germany's hydrogen bus, specifically targeting the year 2050. It features an array of critical technologies, including Electrolyzers, Fuel Cells, the Sabatier process, as well as Hydrogen Tank and Underground Storage. This figure not only showcases these technologies but also integrates original time series data related to battery technology. While these datasets are derived from model outputs, they play as a input role in our research. They form the basis for Fourier Transform (FFT), Continuous Wavelet Transform (CWT), and Cycle Capture (CC) analyses. This integration of technology and data analysis offers a comprehensive understanding of the temporal dynamics, highlighting key trends and variations vital for our study's insights into the future of hydrogen technology in Germany. To access the original time series data for Denmark and Italy in 2050, please refer to Figure~\ref{fig_S2} located in Appendix B.

\subsubsection{Economic Data}
\label{subsubsec:economic-data}

Here we provide techno-economic assumptions for several key components in the energy system. This paper assumes that with technological progress and the expansion of manufacturing scale, the economic costs of renewable energy-related technologies will gradually decrease. The detailed changes are shown in Table~\ref{tab:investment_costs}.

\subsubsection{Storage Data}
\label{subsubsec:storage-data}

When determining key economic indicators like the Levelized Cost of Storage (LCOS), it's crucial to consider factors including the efficiency and lifecycle of the pertinent energy storage technology. Table~\ref{tab:technical_details} provides a detailed overview of parameters associated with this technology. It should be noted that, owing to terrain constraints, Italy was precluded from constructing underground H$_2$ storage in the model.

\begin{table*}[t]
\centering
\caption{Overnight investment cost assumptions per technology and year.}
\label{tab:investment_costs}
\begin{tabular}{lp{1.3cm}p{1.25cm}p{1.25cm}p{1.25cm}p{1.25cm}p{1.25cm}p{1.25cm}p{1.25cm}p{1.1cm}} 
\toprule
Technology & Unit & 2020 & 2025 & 2030 & 2035 & 2040 & 2045 & 2050 & source \\
\midrule
Onshore wind & \euro/kWel & 1118 & 1077 & 1035 & 1006 & 977 & 970 & 963 & \cite{DanishEnergyAgency2019} \\
Offshore wind & \euro/kWel & 2128 & 2031 & 1934 & 1871 & 1808 & 1792 & 1777 & \cite{DanishEnergyAgency2019} \\
Solar PV (utility-scale) & \euro/kWel & 398 & 326 & 254 & 221 & 188 & 169 & 151 & \cite{vartiainenimpact2020} \\
Solar PV (rooftop) & \euro/kWel & 1127 & 955 & 784 & 723 & 661 & 600 & 539 & \cite{vartiainenthetrue2020} \\
Battery storage & \euro/kWh & 232 & 187 & 142 & 118 & 94 & 84 & 75 & \cite{DanishEnergyAgency2019} \\
Battery inverter & \euro/kWel & 270 & 215 & 160 & 130 & 100 & 80 & 60 & \cite{DanishEnergyAgency2019} \\
H$_2$ storage underground & \euro/kWh & 3.0 & 2.5 & 2.0 & 1.8 & 1.5 & 1.4 & 1.2 & \cite{DanishEnergyAgency2019} \\
H$_2$ storage tank & \euro/kWh & 57 & 50 & 44 & 35 & 27 & 24 & 21 & \cite{DanishEnergyAgency2019} \\
Electrolysis & \euro/kWel & 600 & 575 & 550 & 537 & 525 & 512 & 500 & \cite{DanishEnergyAgency2019} \\
Fuel cell & \euro/kWel & 1300 & 1200 & 1100 & 1025 & 950 & 875 & 800 & \cite{DanishEnergyAgency2019} \\
\bottomrule
\end{tabular}
\par\medskip 
\footnotesize
\begin{minipage}{\textwidth}
\textbf{Note:} In reference \cite{vartiainenimpact2020} and \cite{vartiainenthetrue2020}, solar PV investment cost is expressed in 2019-euros. It has been translated into 2015-euros, assuming a growth rate of 2\%. For complete parameters, see Supplementary Table 4 in [25].
\end{minipage}
\end{table*}

\subsection{Code Availability}
\label{subsec:code-availability}

The PyPSA-Eur-Sec-30-path model is available through the repository \url{https://zenodo.org/records/4014807#.X1IKRYtS-Uk}. Code to capture the cycle information in the time series data is available at \url{https://github.com/Zion-tunan/Capture-Cycle-in-Time-Series-Data}. Moreover, code to plot the figures shown in this paper is available at \url{https://github.com/Zion-tunan/Direct-and-Indirect-Hydrogen-Storage-Dynamics-of-InteractionsWithin-Europe-s-Highly-Renewable-Energ}.

\begin{table}[t]
\centering
\caption{Technical details of storage}
\label{tab:technical_details}
\begin{tabular}{lp{0.5cm}p{0.5cm}p{0.5cm}p{0.5cm}p{1.1cm}}
\toprule
Tech & Unit & FOM & Life & Effi. & Source \\
\midrule
Battery storage & kWh & 0.0 & 20 & & \cite{DanishEnergyAgency2019} \\
Battery inverter & kWel & 0.2 & 20 & 0.9 & \cite{DanishEnergyAgency2019} \\
H$_2$ storage underground & kWh & 2.0 & 100 & 1.0 & \cite{DanishEnergyAgency2019} \\
H$_2$ storage tank & kWh & 1.1 & 25 & & \cite{DanishEnergyAgency2019} \\
Electrolysis & kWel & 5.0 & 25 & 0.8 & \cite{DanishEnergyAgency2019, budischakcostminimized2013} \\
Fuel cell & kWel & 5.0 & 10 & 0.58 & \cite{DanishEnergyAgency2019, budischakcostminimized2013} \\
\bottomrule
\end{tabular}
\par\medskip 
\footnotesize
\begin{minipage}{.45\textwidth}
\textbf{Note:} The efficiency and lifetime for battery will also change over time. For complete parameters, see Supplementary Table 5 in [25].
\end{minipage}
\end{table}

\section{Results and Discussion}
\label{results-and-discussion}

\subsection{The Energy System Transition Path in Three Countries}
\label{subsec:the-energy-system-transition-path-in-three-countries}

\begin{figure}[hbtp]
	\centering 
	\includegraphics[width=0.5\textwidth, angle=0]{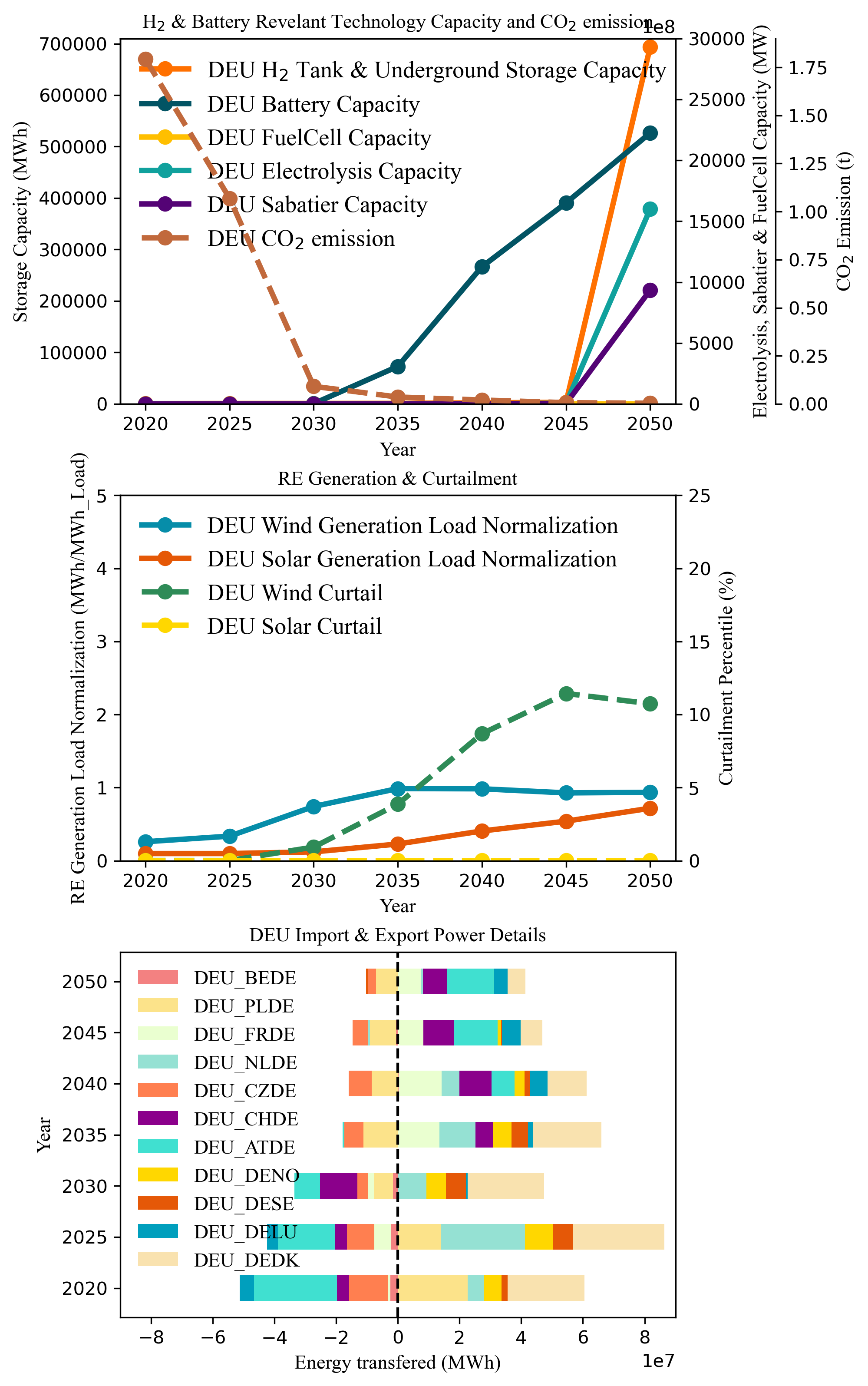}	
	\caption{Evolution of Germany's Energy Transition Landscape: Storage Capacity, CO$_2$ Emission, Renewable Generation \& Curtailment, and Cross-country Flow of Electricity (2020-2050)}
	\label{fig_4}
\end{figure}

In Figure~\ref{fig_4}, we present a comprehensive overview of the energy transition landscapes in Germany from 2020 to 2050, and display the energy transition landscapes in Denmark and Italy in Figure~\ref{fig_S3}. The two figures detail the evolution of storage capacity, CO$_2$ emissions, renewable energy generation \& curtailment and cross-country flow of electricity across the three countries, offering a comparative insight into their respective energy transitions. Please note that regarding cross-country flow of electricity, positive values represent energy import from other countries, and negative values represent export.

The first subgraph of Figure~\ref{fig_4} carefully examines Germany's carbon emission reductions alongside changes in energy storage capacity over the years. It reveals that energy storage played a minimal role in Germany's initial rapid decarbonization phase but became increasingly significant in the later stages of deep decarbonization. Notably, battery involvement in decarbonization emerges around 2035. The capacity of electrolyzers, Sabatier processes, H$_2$ tanks, and underground storage sees a substantial growth by 2050, indicating large-scale involvement in H$_2$ storage. Concurrently, the capacities of H$_2$ tanks and underground storage experience significant growth. However, fuel cell capacity remains negligible, suggesting that direct H$_2$ storage predominantly occurs through the Sabatier process.

The second subgraph illustrates Germany's significant increase in wind and solar power generation capacity, crucial for achieving decarbonization goals. Up to 2030, wind power serves as the primary decarbonization driver, increasing by $121\%$ compared to 2025. Post-2035, solar power becomes more important, maintaining a steady average annual growth rate of $58\%$ until 2050. This period sees a concurrent rise in both solar power generation and battery capacity, underlining the growing renewable energy penetration in Germany.

In the context of cross-country flow of electricity, Germany initially appears as a net electricity importer but also contributes significantly to exports. In 2020, exports accounted for $11\%$ of its total electricity consumption. However, there's a year-on-year decline in Germany's electricity exports, indicating a decreasing role in Germany international power transmission as a storage mechanism.

Denmark, a wind-dominated country, shows a markedly smaller battery storage capacity than Germany, projected to reach 8 GWh by 2050. Both direct and indirect H$_2$ storage capacities in Denmark, except for fuel cells, are set to increase substantially from 2045, five years earlier than in Germany. Notably, Denmark, as a net electricity exporter, sees its output comprising $74\%$ of its total electricity load from 2020 to 2050. In contrast, Italy, a country where solar energy is more prevalent, experiences irregular yearly variations in electricity transmission with other countries. Italy primarily depends on battery and indirect H$_2$ storage, opting forgoes direct H$_2$ storage. Interestingly, Italy's capacity of electrolyzers and Sabatier processes significantly exceeds that of Germany.

In summary, this section provides basic insights into the energy transition pathways of Denmark, Italy, and Germany up to 2050. A key similarity across these countries is the increasing penetration rate of renewable energy and the collective move away from fuel cells. Another important commonality is the early-stage reliance on cross-country electricity flow, playing a pivotal role during the initial rapid decarbonization phase, before significant energy storage involvement. As the transition progresses, however, their approaches begin to diverge. Denmark shows a smaller battery storage capacity compared to Germany, yet both nations are significantly ramping up their 2 kinds of hydrogen storage capacities, with Denmark starting this increase five years earlier. Denmark also stands out as a major electricity exporter, with a substantial part of its output derived from renewable sources. Italy exhibits a different pattern, with a different in the variation of electricity transmission with other countries and a greater reliance on battery and indirect hydrogen storage solutions.

Regarding the scope of our subsequent analysis, we will narrow our focus to specific time frames for each type of energy storage based on the conclusion from this section. For battery storage, the analysis will concentrate on the period from 2030 to 2050. This timeframe is crucial as it marks the phase when battery storage is expected to play a more prominent role in the energy transition. For H$_2$ storage, our examination will be confined to the decade from 2040 to 2050, a critical period for the maturation and increased implementation of hydrogen-based energy solutions.

\subsection{Quantitative Analysis: Economic Performance of Storage}
\label{subsec:quantitative-analysis-Economic-Performance-of 
-Storage}

\begin{figure}[tbhp]
	\centering 
	\includegraphics[width=0.45\textwidth, angle=0]{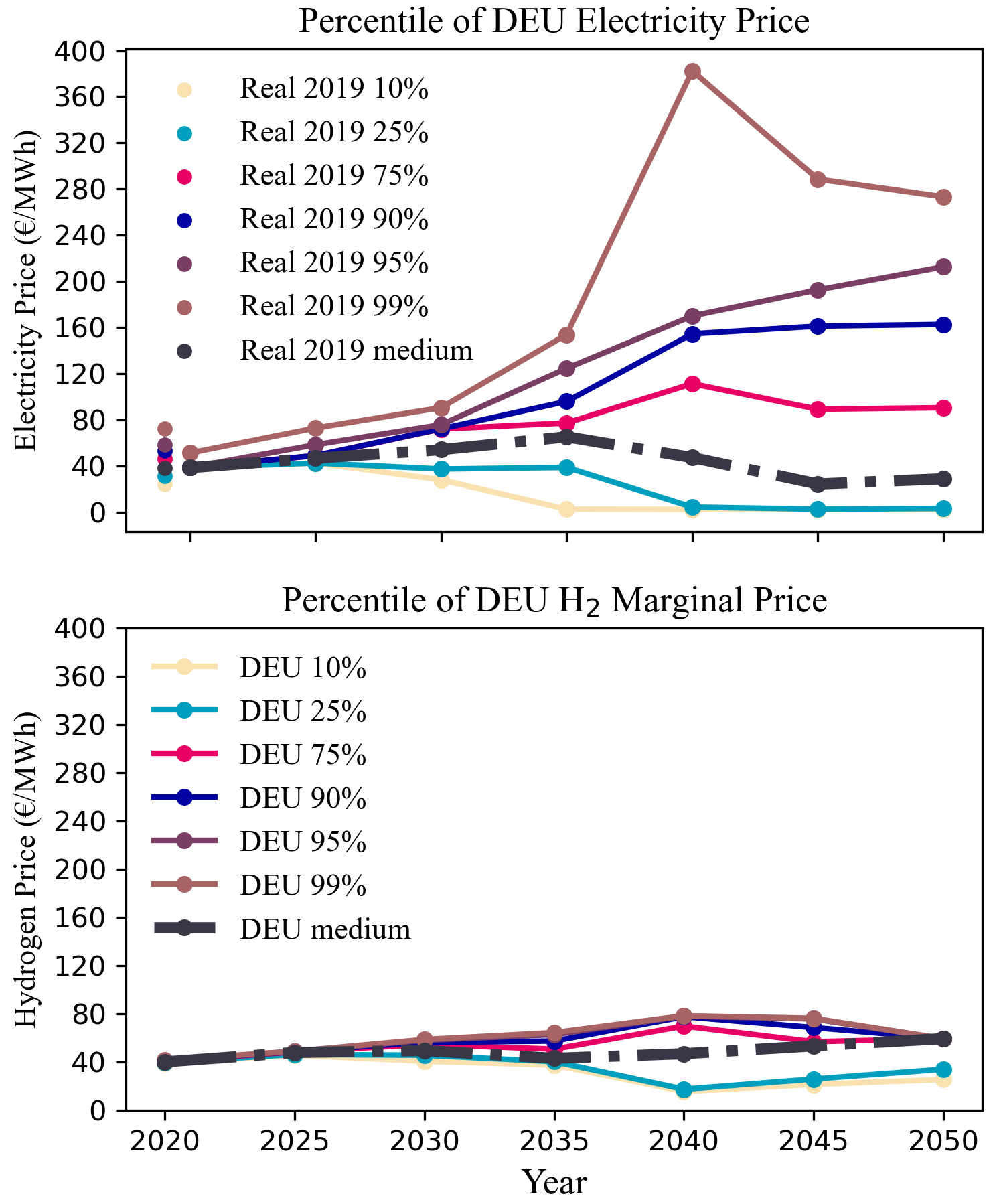}	
	\caption{Historical and Projected Electricity Price and H$_2$ Marginal Price Distribution in Germany: A 2050 Outlook}
	\label{fig_5}
\end{figure}

In Figure~\ref{fig_5}, we illustrate the temporal evolution of electricity price percentiles and H$_2$ marginal price percentiles for Germany. For an extended comparative analysis involving Denmark and Italy, and for the distribution patterns of electricity and H$_2$ marginal prices across these nations, we refer the readers to Appendix B supplementary figures~\ref{fig_S4}, \ref{fig_S5}, and \ref{fig_S6}. The analysis of the electricity and H$_2$ price spread can provide a basis and reference for subsequent analysis of the energy storage benefits.

Figure~\ref{fig_5} and Figure~\ref{fig_S4} indicates a common pattern among the three countries concerning electricity prices: a spreading of the price, denoted by the disparity between peak and trough values, which escalates progressively until 2040 before exhibiting a contraction. The median volatility of electricity prices maintains relative stability, with average fluctuations confined to below 53\% for the countries in question. This stability, however, is not indicative of a clustering around the median price. Instead, it reflects a balance between the increasing occurrences of both high and low price extremes, as evidenced directly in Figure~\ref{fig_S4}. Here, we observe a transition in electricity prices from a tight band of 25 €/MWh to 50 €/MWh to a broader distribution prior to 2040, subsequently reconverging but remaining more varied than in the baseline year of 2020. From 2020 to 2035, the disparity in electricity prices among Denmark, Germany, and Italy exhibited a remarkable similarity. However, future projections suggest a notable divergence in this trend. It is expected that from 2040, the electricity price differential in Denmark and Germany will converge and significantly exceed that of Italy. Specifically, the price spread in Denmark and Germany is projected to be 30\% higher than in Italy during that year.

The spread of H$_2$ marginal prices is primarily influenced by fluctuations in electricity prices, given that hydrogen production is reliant solely on electrolysis. Notably, the spread of H$_2$ marginal prices exhibits an initial increase, peaking in 2040, before showing a decline. This trend is substantiated by Figure~\ref{fig_S5}, which illustrates the H$_2$ marginal price shift from a narrow range of 20 €/MWh to 40 €/MWh to a broader distribution prior to 2040, followed by a convergence post-2040. While the distribution trends of H$_2$ marginal prices align with those of electricity prices, a notable shift in the relative positioning of H$_2$ marginal price spreads among the three countries has been observed. Contrasting with the behavior of electricity price spreads, projections from 2045 onwards indicate a reordering in the H$_2$ marginal price spreads. This new arrangement places Italy at the highest level, followed by Germany, and then Denmark, representing a significant deviation from electricity price spread trends. This inconsistency underscores that while electricity prices significantly influence H$_2$ marginal prices, they do not exclusively determine all facets of H$_2$ marginal prices.

\begin{figure}[tbhp]
	\centering 
	\includegraphics[width=0.48\textwidth, angle=0]{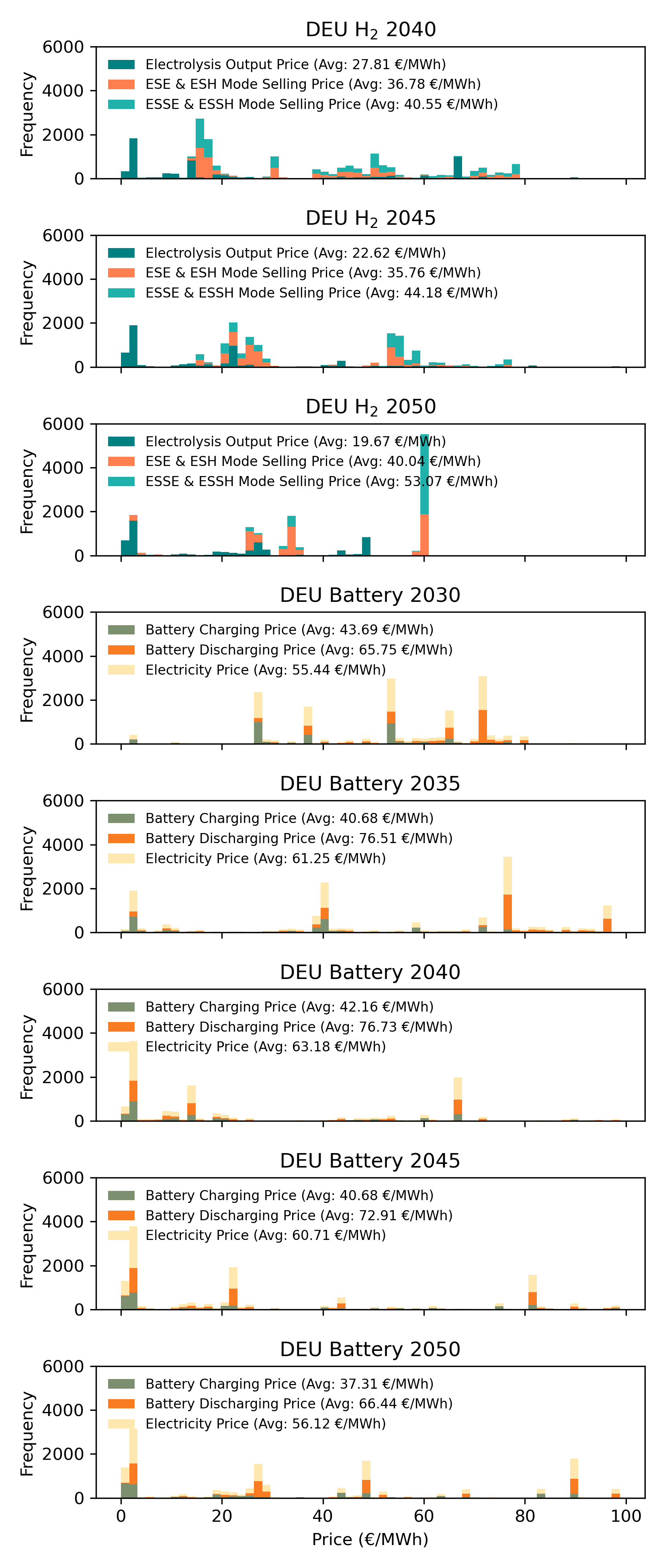}	
	\caption{Comparative Price Distribution for Battery Storage Buying and Selling and Electricity (2030-2050) with Detailed Hydrogen Buying and Selling Analysis (2040-2050) in Germany}
	\label{fig_6}
\end{figure}

In Figure~\ref{fig_6}, we present the dynamics of Germany storage buying and selling prices alongside electricity price distributions. In this analysis, the terms 'buying price' and 'selling price' refer to the electricity or H$_2$ prices at the specific times when the storage buys or sells energy. Therefore, the objective of Figure~\ref{fig_6} is to examine the interplay between the buying and selling prices of energy storage, and the prices of electricity and H$_2$. Notably, the buying price for H$_2$ storage corresponds to the electricity price for powering electrolyzers, while the selling price is linked to the marginal price of H$_2$ as used in the Sabatier process. It's important to distinguish between two hydrogen storage approaches: direct and indirect. Direct H$_2$ storage, as depicted here, excludes fuel cells, based on their infeasibility for energy transition demonstrated in Section~\ref{subsec:the-energy-system-transition-path-in-three-countries}. In contrast, for battery storage, the buying and selling prices directly reflect current electricity market rates. Figure~\ref{fig_6} also integrates the battery buying and selling prices with the broader distribution of electricity prices. For detailed comparisons of storage pricing in Denmark and Italy, refer to Figure~\ref{fig_S7} in Appendix B.

In Figure~\ref{fig_6}, our observation reveals that in Germany, the buying price for energy from storage systems is typically lower, while the selling price is higher. The average selling price is lowest for energy selling from the indirect hydrogen storage, moderate for the direct hydrogen storage, and highest for battery storage. This pattern suggests the tank or underground H$_2$ storage from the direct hydrogen storage works well. Simply put, energy storage systems earn revenue by acquiring energy at a lower buying price and selling it at a higher selling price. Therefore, the gap between the lower energy prices and the higher energy prices becomes the main profit margin for energy storage. The difference in buying and selling prices across the three types of energy storage is in line with the trends in electricity and H$_2$ marginal price spreads observed in Figure~\ref{fig_5}, where the range of electricity price variation exceeds that of H$_2$ marginal price.

However, the largest average price difference among the three energy storage methods is seen in 2050. For indirect hydrogen storage, the maximum difference is 18 €/MWh, for direct hydrogen storage it's 24 €/MWh, and for battery storage, it reaches 38 €/MWh. This contrasts with the observation in Figure~\ref{fig_5}, where the peak spread difference between the two energy carrier is noted in 2040. In the cases of Denmark and Italy, as shown in Figure~\ref{fig_S7}, a similar trend to Germany is observed. This indicates that energy prices are not the sole determinant of energy storage profit margins, and a more detailed investigation into the operational aspects of energy storage is warranted.

\begin{figure}[tbhp]
	\centering 
	\includegraphics[width=0.4\textwidth, angle=0]{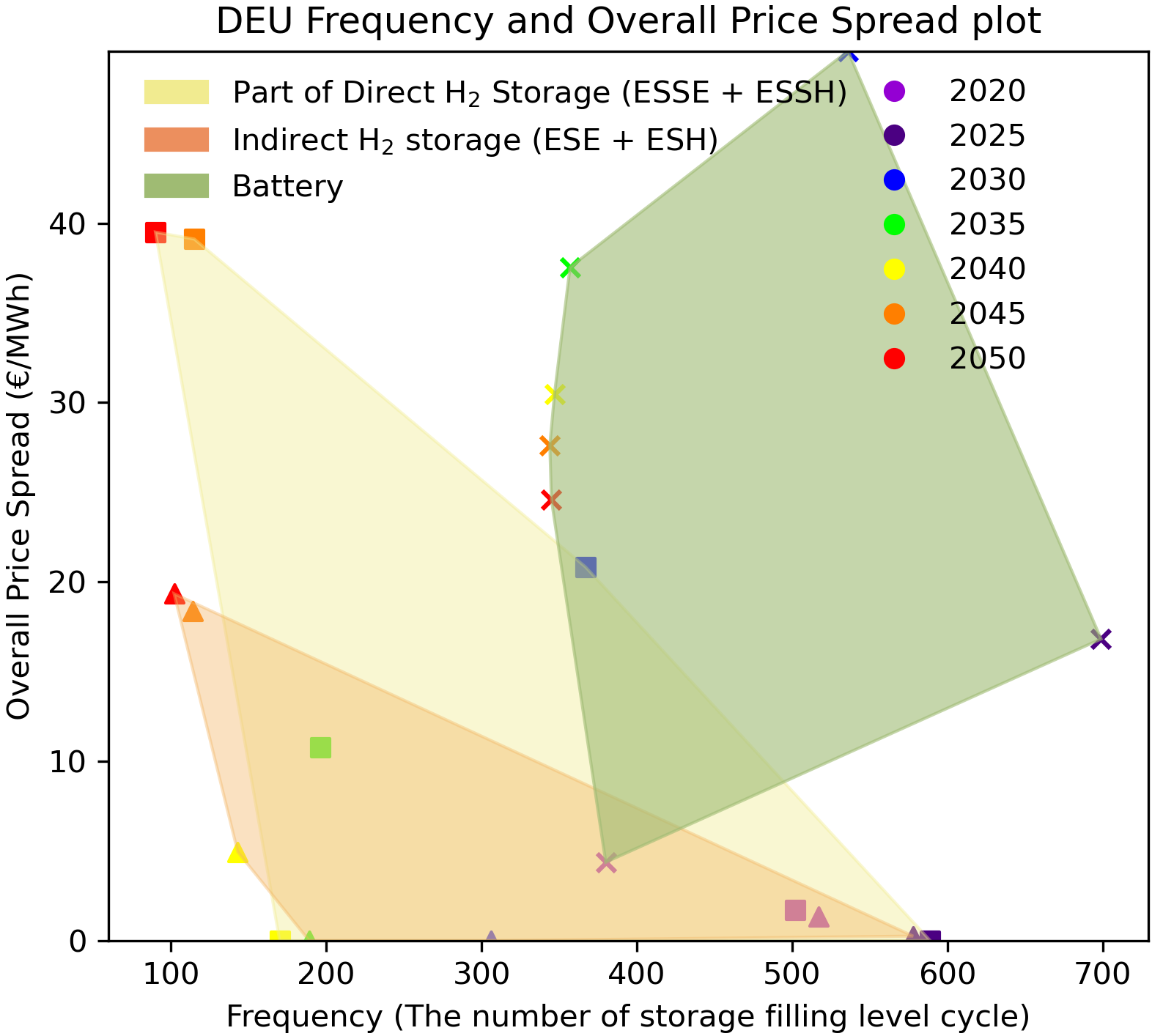}	
	\caption{Analysis of Cycle Frequency and Overall Price Spreads in German H$_2$ Storage and Battery Systems (2020-2050)}
	\label{fig_7}
\end{figure}

Figure~\ref{fig_7} presents the frequency (number of cycles) and overall price spread (OPS) of two types of hydrogen storage in Germany (excluding ESFC working modes from direct H2 storage due to fuel cell technology is unreliability under this study's assumptions) along with battery storage. The frequency is determined using the Cycle Capture (CC) Method. The OPS, also understood as the cycle-cased profitability index, quantifies the average difference in selling and buying prices within identified storage filling cycles over the year. For an in-depth explanation of OPS, see Equation~\ref{equation_3}. Thus, Figure~\ref{fig_7} offers a detailed view of the operational dynamics of energy storage within the energy system, shedding light on the link between the profitability of energy storage, electricity prices, hydrogen price and the actual operation of energy storage. For a comparative analysis of hydrogen and battery storage performance in Denmark and Italy, refer to Figure~\ref{fig_S8} in Appendix B.

From Figure~\ref{fig_7}, we note that in the year marking Germany's shift to energy storage, the cycle frequencies for both indirect and direct H$_2$ storage were roughly 100 times per year, significantly lower than the battery storage cycle frequency, which was about 345 times per year. However, direct H$_2$ storage demonstrated the highest OPS, followed by battery storage, and then indirect H$_2$ storage. As shown in Figure~\ref{fig_S8}, while the relative cycle frequencies of these energy storages remain consistent across the three countries, their OPS values vary in Denmark and Italy. In Denmark, direct H$_2$ storage's OPS is comparable to that of battery storage. In Italy, the OPS for indirect H$_2$ storage is similar to that of battery storage. It's also notable that the cycle frequencies for these three types of energy storage show considerable variation among the three countries. For direct and indirect H$_2$ storage, the frequency order from highest to lowest is Denmark, Germany, and Italy. For battery storage, the frequency order is reversed, with Denmark being the lowest and Italy the highest. This observation underscores how the actual operational practices of energy storage systems significantly affect their profitability, particularly in terms of storage charging and discharging frequencies and the associated price differences.

\begin{figure*}[t]
	\centering 
	\includegraphics[width=0.9\textwidth, angle=0]{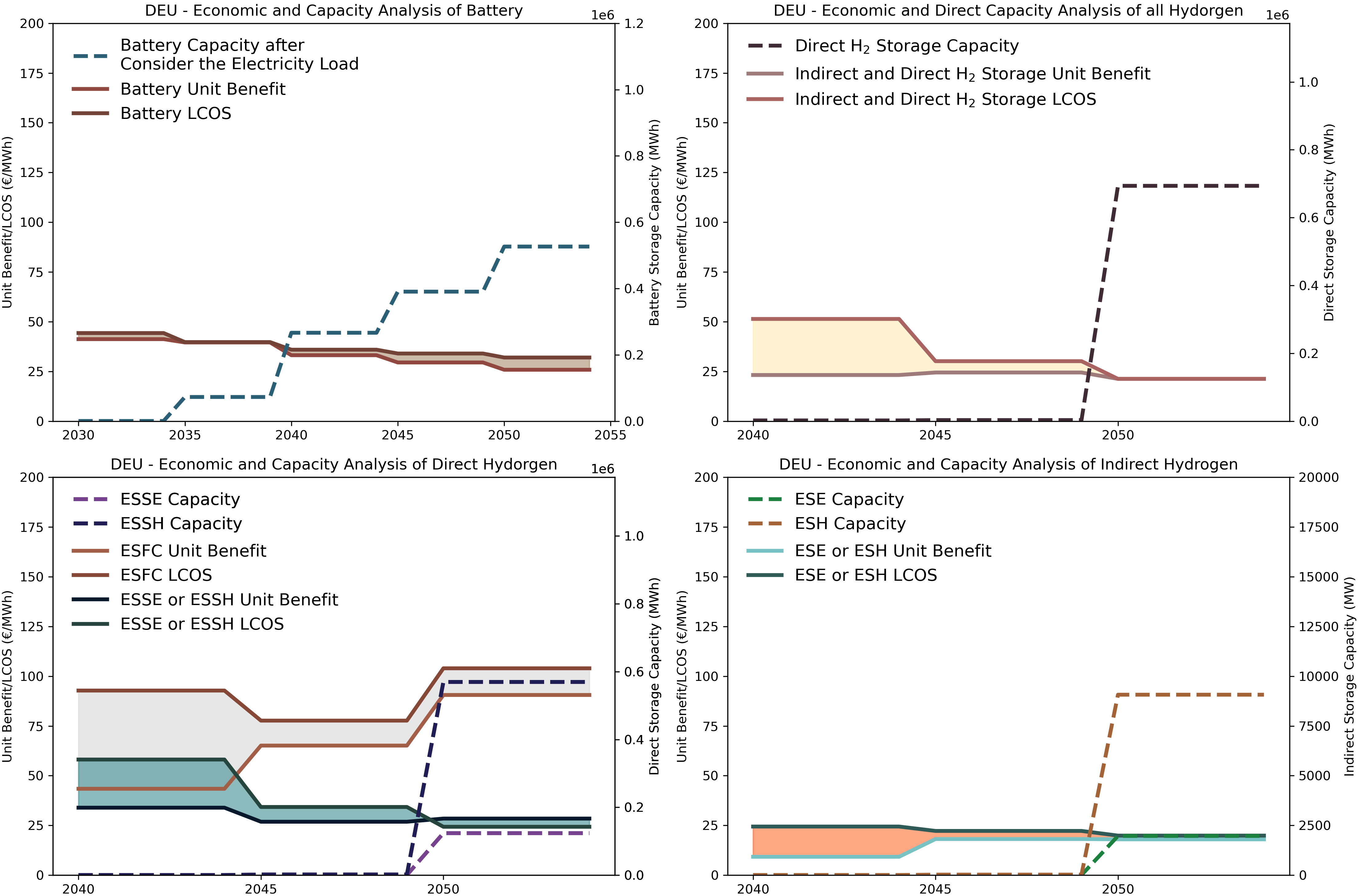}	
	\caption{Temporal Evolution of LCOS, Unit Benefits, and Capacities of Energy storage in Germany in the Late Energy Transition Stage}
	\label{fig_8}
\end{figure*}

Figure~\ref{fig_8} indicates the evolution of the cumulative levelized cost of storage (LCOS), unit benefit, and the installed capacities of hydrogen and battery storage systems in Germany. Each subplot is structured with dual y-axes: the left axis quantifies economic parameters, while the right axis measures the capacities of energy storage technologies. The initial subgraph of the analysis focuses on the German battery performance. Subsequently, the second subgraph examines all H$_2$ storage economic performance and direct H$_2$ storage capacity. The third subgraph addresses the economic performance and capacities of three hydrogen working modes associated with direct H$_2$ storage: Electrolysis - Store - Fuel Cell (ESFC), Electrolysis - Store - Sabatier - Electricity (ESSE), and Electrolysis - Store - Sabatier - Heating (ESSH). It is important to note that our model limited that the calculations for ESSE and ESSH modes cannot be separated on hour level but rather on an annual basis. Consequently, the LCOS and unit benefits for ESSE and ESSH are identical. A similar situation arises with the two indirect H$_2$ storage working modes: Electrolysis - Sabatier - Store - Electricity (ESE) and Electrolysis - Sabatier - Store - Heating (ESH), as demonstrated in the fourth subplot of our study. Notably, the measurement of indirect H$_2$ storage capacity corresponds to the segment of the electrolyzer that supply energy directly to the Sabatier process, denominated in megawatts (MW). Conversely, the capacity of direct H$_2$ storage is measured by the volume of hydrogen that can be stored underground or within tanks, denominated in megawatt-hours (MWh).

To contextualize these findings within a broader European framework, Figure~\ref{fig_S9} offers a comprehensive comparison of these indicators across Denmark, Germany, and Italy. Complementarily, Figure~\ref{fig_S10} sheds light on the non-cumulative levelized cost of storage for the three countries. The term 'non-cumulative' herein implies that at each time step, the costs are appraised based on the current price levels, ignoring the integration of historical costs. In this model, the point where the LCOS and unit benefit meet in Figure~\ref{fig_S10} is important because it shows when the money spent on the technology has been made back, which could also indicate when we might expect to see more investment in the technology's capacity. This point highlights a limitation in the model we used for this study. To show the real costs involved in moving to new energy technologies, Figure~\ref{fig_8} includes the past costs that have built up over time. Additionally, we've added Figure~\ref{fig_S10} in Appendix B to check that our findings are accurate.

In Germany, the convergence of the LCOS and the unit benefit for battery storage is projected to occur in 2035. This indicates that, from 2035 onwards, battery technology will start to provide economic value and its capacity is expected to increase substantially. However, a downward trend in unit benefits is detected from 2035 to 2050, with a decrease from the initial 40 €/MWh to 26 €/MWh. This trend deeper underscores the disconnect between the changes in energy storage economic performance and the fluctuations in electricity price spreads.

Regarding H$_2$ storage in Germany, it is not until 2050 that we foresee the LCOS aligning with the unit benefit, with a significant expansion in its capacity. The projected unit benefit for H$_2$ storage in 2050 is 21 €/MWh. When comparing the unit benefits of battery and H$_2$ storage, it's evident that although the former is higher, it also implies a greater investment per unit of energy. Focusing specifically on direct and indirect H$_2$ storage, their respective unit benefits in 2050 are anticipated to be 28.5 €/MWh and 18 €/MWh. It is also notable that most of the energy stored in both these forms of H$_2$ storage is likely to be towards the heating sector. This is evident from the fact that the capacities for ESSH and ESH are projected to be higher than those for ESSE and ESE, respectively.

In examining battery storage growth in Denmark, Germany, and Italy, we observe a simultaneous increase across these countries. By 2035, the unit benefits of battery storage are expected to outweigh their LCOS. Denmark exhibits the highest unit benefit and LCOS, followed by Germany, and then Italy with the lowest. This variation likely results from Italy's higher frequency of battery usage compared to Denmark’s lower usage frequency, as evidenced in Figure~\ref{fig_7} and Figure~\ref{fig_S21}. Regarding storage capacity, Denmark, Germany, and Italy follow an ascending order. Remarkably, when comparing absolute values of battery storage capacity, Denmark's figures appear less significant, especially when shown alongside Germany's in Figure~\ref{fig_S9} and Figure~\ref{fig_S10}. However, this comparison doesn't account for Denmark's substantially smaller electricity load, which is 16 times less than Germany's. By multiplying Denmark's capacity figures by 16, we can more accurately appreciate the extent of its expansion, highlighting the critical role of battery storage in a country heavily reliant on wind energy like Denmark.

In contrast, the dynamics of H$_2$ storage in these countries present a different picture. Denmark shows the lowest unit benefit and LCOS for H$_2$ storage, with Germany highest and Italy in between. This suggests that Italy finds building H$_2$ storage more appealing compared to Germany. Italy's preference for indirect H$_2$ storage over direct H$_2$ storage, due to the latter’s inability to recover initial investments, is also notable. After accounting for the influence of electricity load, Italy's direct H$_2$ storage capacity still surpasses that of Germany. What’s more, the ESFC plans in all three countries indicate that they are unlikely to recoup initial investments by 2050, explaining the lack of acceptance for these plans under this study’s assumptions.

\begin{figure}[t]
	\centering 
	\includegraphics[width=0.2\textwidth, angle=0]{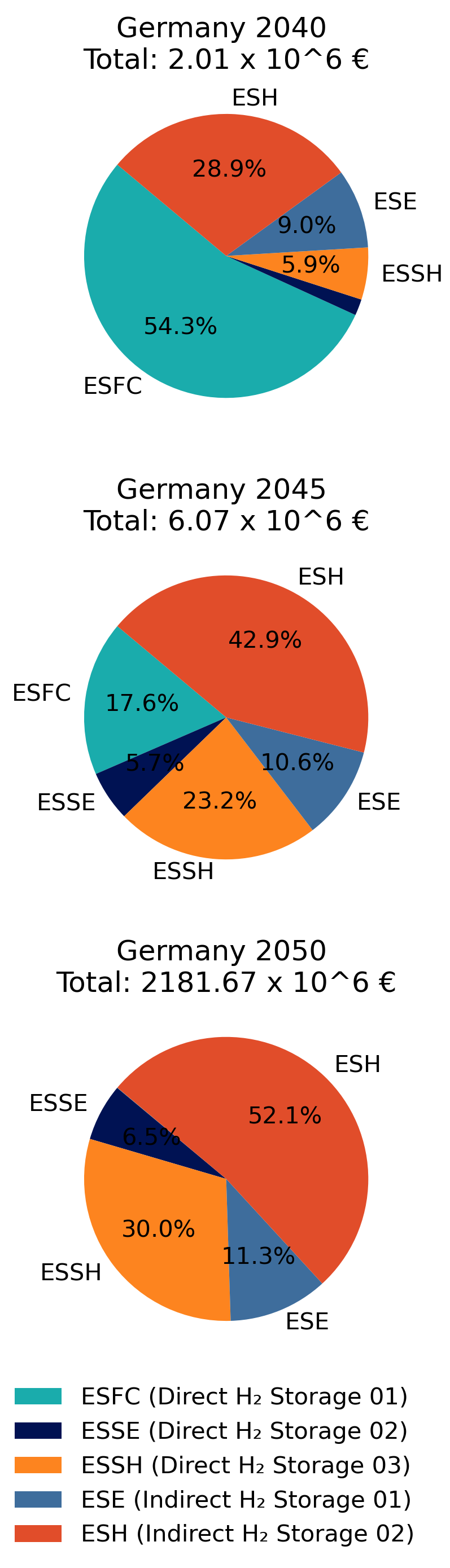}	
	\caption{Revenue Distribution of Germany's H$_2$ Storage Working Modes in 2040, 2045, and 2050}
	\label{fig_9}
\end{figure}

Figure~\ref{fig_9} presents the revenue distribution of five hydrogen storage modes in Germany for the years 2040, 2045, and 2050. Focusing on the total revenue, we note a significant increase by 2050, reaching €2.2 billion. This figure is more than $99.7\%$ higher than the total revenue in 2045, and even more so compared to 2040. In the year 2050, indirect H$_2$ storage contributes to $63.4\%$ of Germany’s total H$_2$ storage revenue, indicating its predominant role in revenue generation. Additionally, a notable aspect is that the majority of H$_2$, whether from indirect or direct storage, is used in the heating sector, accounting for $82.1\%$ of the revenue.

Similar patterns are observed in Denmark and Italy, as shown in Figure~\ref{fig_S11}. Italy relies exclusively on indirect H$_2$ storage, while in Denmark, indirect H$_2$ storage comprises $75.3\%$ of the total H$_2$ storage revenue. Remarkably, in 2050, Italy leads the three countries with a total revenue of €3.3 billion, followed by Denmark with €2.4 billion.

\subsection{Qualitative Analysis: Time – Frequency Interdependency Analysis}
\label{subsec:qualitative-analysis: time–frequency-interdependency-analysis}

\begin{figure*}[t]
	\centering
	\includegraphics[width=0.9\textwidth, angle=0]{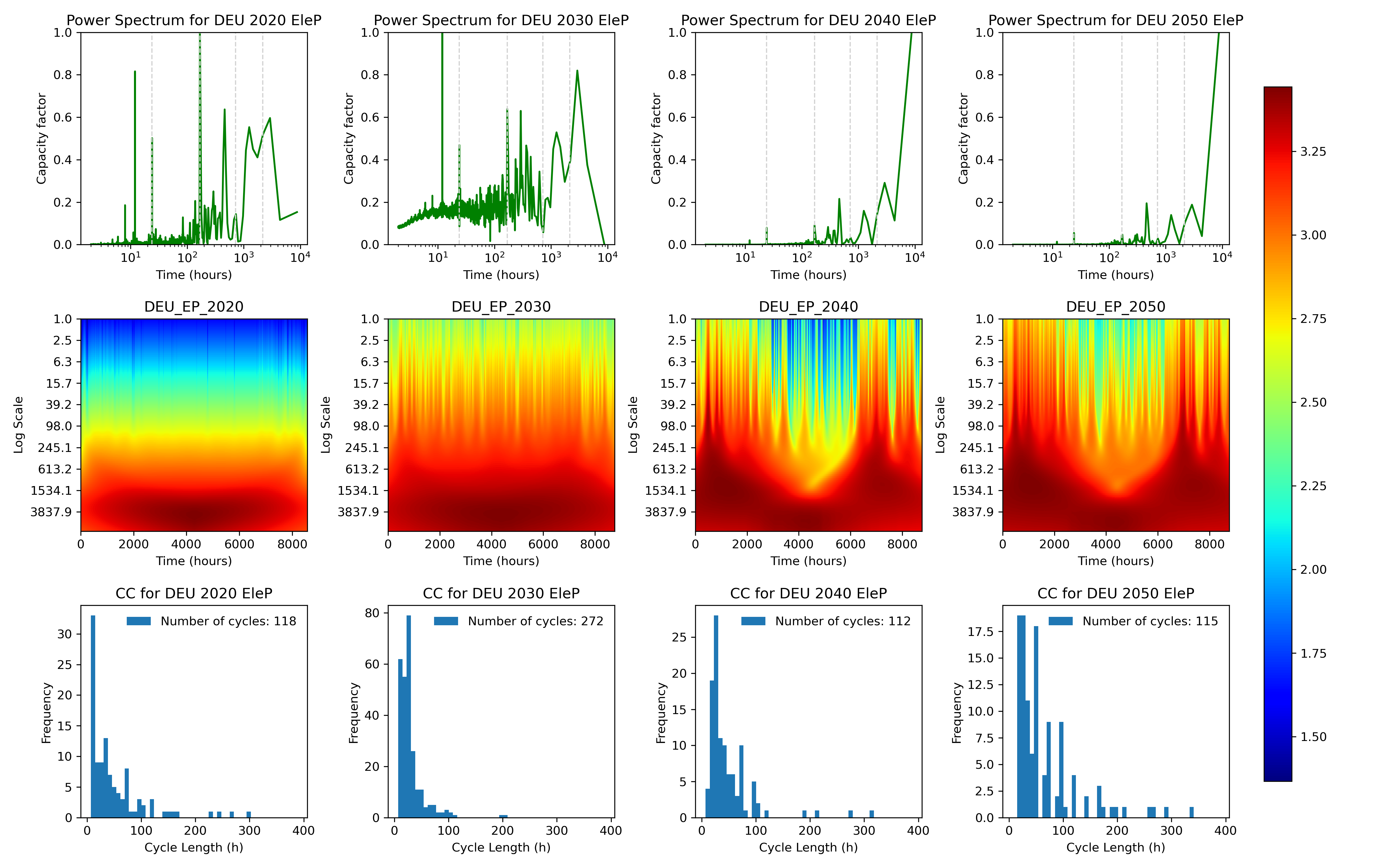}	
	\caption{Wavelets, Cycles, and Spectra: Deciphering Germany’s Electricity Price Evolution in 2020, 2030, 2040 and 2050}
	\label{fig_10}
\end{figure*}

In this section, we use Fast Fourier Transform (FFT), Continuous Wavelet Transform (CWT), and Cycle Capture (CC) methods to study changes in electricity prices, renewable energy generation, and energy storage filling levels in three different countries. Our study examines snapshot data at seven points, every five years from 2020 to 2050. The selected methodologies aim to accurately detect genuine fluctuations within the collected data, facilitating a more in-depth validation of the interaction among different segments of the energy system as inferred from these fluctuations. The variation in results from different volatility analysis methods does not imply inaccuracies in any method, rather, it highlights the unique focal points each method offers. And here, we're only showing Germany's electricity price fluctuation figure for 2020, 2030, 2040, and 2050 in the main body of the article, as shown in Figure~\ref{fig_10}. 

The ~\ref{appendix b}, specifically Figures ~\ref{fig_S12} to ~\ref{fig_S26}, provides detailed data on the changes in electricity prices, renewable energy production, and levels of energy storage from 2020 to 2050 across three countries. This section will first examine the fluctuations in these areas separately, highlighting the real trends in each dataset. Then, we'll connect these various pieces of information to understand the relationship among electricity prices, renewable energy generation and energy storage. It is worth to note that in the Cycle Capture method, we set the upper and lower boundaries for both renewable energy power generation and storage filling levels at $10\%$of their normalized values. For the electricity price, the limit was set at $5\%$.

Figure~\ref{fig_10} presents varied analytical outcomes on electricity prices in Germany across four distinct years, derived from three different methods. The variation in results from different volatility analysis methods does not imply inaccuracies in any method, rather, it highlights the unique focal points each method offers. This section will first synthesize insights from all three methods to unveil the authentic fluctuations in the assorted data sets. Subsequently, this paper will interconnect these disparate data to create a link between electricity prices and the energy storage and renewable generation that have been identified as key factors influencing electricity prices.

For the FFT analysis results of electricity prices, as displayed in Figure~\ref{fig_10} and Figure~\ref{fig_S13}. Denmark's electricity prices fluctuated mainly on larger scales such as weekly, monthly, and seasonally. In Germany, fluctuations ranged widely from daily to seasonal scales. Italy saw changes mostly daily and seasonally. However, from 2040, price fluctuations became more evenly spread across all scales. The intensity of fluctuations at nearly all scales decreased, with Italy experiencing the most significant reduction, showing only minor dominant fluctuations daily and seasonally. Denmark was the least affected by this decrease.

The CWT depicted in Figure~\ref{fig_10} and Figure~\ref{fig_S12} elucidates the temporal evolution of frequency components within the electricity prices for Denmark, Germany, and Italy. Commencing in 2040, robust seasonality is discernible across all three nations, aligning with the observations made via FFT. This seasonality is particularly pronounced on the seasonal scale from 2040 onwards. CWT, however, offers a more nuanced insight, revealing that the apparent surge in seasonality during this period is attributed to the attenuation of frequencies on shorter time scales during the summer months. Conversely, during the winter months, the frequency components on these shorter time scales are retained. A peculiar observation is noted for Italy, where the strength of seasonality peaks in 2040 and then experiences a gradual decline. The salience of frequencies across most scales seems to harmonize over time, yet distinctions in the seasonal and diurnal scales remain perceptible. And CWT results further proves what was said at the previous paragraph that the information or variability in the electricity prices is more evenly distributed across different frequency scales.

The CC analysis results, as presented in Figure~\ref{fig_10} and Figure~\ref{fig_S14}, reveal variations exceeding $5\%$ in the electricity prices across three countries. First, these variations are predominantly observed on daily and weekly scales. This suggests that despite our focus on the larger-scale frequency components in the FFT and CWT, the short-term frequency components remain crucial throughout. They offer insights into the intricate dynamics of the energy system. Second, the annual trend of these fluctuations aligns with the annual variations in the FFT frequency component on daily and weekly scales. Predictions suggest that the frequency component of electricity prices in these countries, based on daily and weekly scales, will peak in 2030. This indicates the appropriateness of the parameter settings for electricity price changes in the CC analysis.

In summary the analysis of the electricity price across three countries, the patterns of frequency fluctuations in electricity prices vary among the three countries. Denmark primarily focuses on large time scale fluctuations, Germany considers fluctuations across all scales, while Italy mainly highlights those on a daily scale. The daily significance frequency of electricity prices in all three nations reaches its zenith in 2030. By 2040, frequencies start to spread uniformly across scales compare to before. This paper subsequently carried out same analysis for energy storage filling level and renewable energy generation using.

Next, we enter the analysis of energy storage filling level. Readers need to refer to Figures ~\ref{fig_S15} to ~\ref{fig_S21} in ~\ref{appendix b}. The battery filling levels, as depicted in Figures ~\ref{fig_S15}, ~\ref{fig_S20}, and ~\ref{fig_S21}, predominantly exhibit fluctuations on daily scale across all three examined countries. Notably, Denmark and Germany also exhibit fluctuations on a weekly scale, with Denmark showing a stronger tendency. Similar to electricity prices, daily scale fluctuations peak in 2030 as indicated by the CC results.

Distinct frequency characteristics are observed in the direct and indirect H$_2$ storage fluctuations across different countries, as delineated in Figures ~\ref{fig_S15} to ~\ref{fig_S19}. Focusing on the direct H$_2$ storage fluctuations, we find that Italy, as our previous analysis, lacks direct H$_2$ storage capacity along its transition path. Therefore, our attention is directed towards Denmark and Germany. In these countries, the primary frequencies are predominantly around the monthly scale. The CC results show that charge and discharge energy fluctuations exceeding $10\%$ still occur on a daily or weekly scale. Furthermore, the CWT analysis suggests that while clear seasonality is absent in the both countries, the frequency of direct H$_2$ storage on the short-period scale is gaining increasing prominence.

Regarding the performance of indirect H$_2$ storage fluctuations, a distinct different pattern emerges when compared to the direct H$_2$ storage. Interestingly, in the years when indirect H$_2$ storage became significant, its main fluctuation pattern matched exactly with that of electricity prices in size and intensity. Moreover, during summer, indirect H$_2$ storage displayed strong daily fluctuations, perfectly complementary with the changes in electricity prices. This indicates that indirect H$_2$ storage is effectively adjusting to the varying energy structures of different countries. 

For renewable energy generation, Figures ~\ref{fig_S22} to ~\ref{fig_S26} highlight that wind energy mainly fluctuates weekly, monthly, and seasonally, while solar energy fluctuates daily and seasonally. Before energy storage came into play, the fluctuations in electricity prices closely matched the patterns of renewable energy generation, significantly influencing price variability.

This section examines the volatility of electricity prices, renewable energy production, and energy storage levels using three methods, uncovering some possible connections. It's important to consider these findings alongside those in section~\ref{subsec:the-energy-system-transition-path-in-three-countries}. This paper use the moment energy storage starts impacting the energy system as a key point to explore changes in electricity prices. Before 2040, fluctuations in electricity prices are largely driven by renewable energy volatility. Afterward, prices are influenced by how well renewable energy and energy storage balance out. Notably, the relationship between electricity prices and energy storage is mutual. While we only focus on how energy storage affects electricity prices in this section, it's interesting to see that indirect H$_2$ storage adjusts its impact according to different energy systems, showing varying patterns of fluctuation. In contrast, battery storage and direct H$_2$ storage influence electricity prices on short-duration and long-duration bases, respectively.

\section{Conclusions}
\label{conclusion}

Energy storage, renewable energy generation, and electricity prices in highly renewable Europe have been investigated through a network of one node per country, which is resolved every hour of the European energy system and is aimed to achieve a $95\%$ carbon reduction by 2050. From the data of 30 countries, three countries – Denmark, Germany, and Italy – with distinct climates, were chosen for a deeper examination. This article aims to elucidate the intricate dynamics within the energy system. To this end, it outlines five distinct H$_2$ working methodologies, classifying them into two categories: direct H$_2$ storage and indirect H$_2$ storage. This paper delves into the energy system's complexities from both quantitative and qualitative perspectives.

The article outlines specific strategies for the energy transition in Denmark, Germany, and Italy, setting the stage for our analysis. Initially, energy storage didn't contribute to the quick reduction of carbon emissions in these countries, with international electricity transmission adding flexibility to their energy systems. Fuel cell technology was set aside by all three countries due to cost concerns—it needs to be more affordable to be considered a feasible option for the energy transition. Renewable energy's impact on the energy system was fixed, shaping electricity price fluctuations before the introduction of energy storage. Once energy storage came into play, electricity prices began to reflect the balance between renewable energy and the different types of energy storage. 

It's important to note the two-way relationship between electricity prices and energy storage. From the perspective of how electricity prices affect energy storage, the spread of prices is a major source of profit for energy storage solutions. However, the profitability of energy storage also depends on the specific energy structure of a location, which influences how it operates. On the other hand, looking at how energy storage affects electricity prices, each of the three energy storage methods uniquely reduces price fluctuations by leveraging different patterns. Among these, indirect H$_2$ storage stands out as a common choice in all three countries, thanks to its better economic performance and its ability to adapt flexibly to various energy structures. In Denmark, battery storage faces capacity constraints, and Italy has moved away from direct H$_2$ storage. These developments further underline the crucial role of indirect H$_2$ storage in advancing the transition to cleaner energy.

\section*{CRediT authorship contribution statement}
\label{CRediT-authorship-contribution-statement}

\noindent \textbf{Zhiyuan Xie}: Writing – review \& editing, Writing – original draft, Visualization, Validation, Coding, Methodology, Investigation, Formal analysis, Conceptualization. \textbf{Gorm Bruun Andresen}: Writing – review \& editing, Supervision, Methodology, Funding acquisition, Conceptualization.

\section*{Declaration of generative Ai in scientific writing}
\label{declaration-of-generative-Ai-in-scientific-writing}
\noindent During the preparation of this work, the authors used OpenAI in order to generate some icons for Figure 1. After using this tool, the authors reviewed and edited the content as needed and takes full responsibility for the content of the publication.

\section*{Acknowledgements}
\label{acknowledgements}
\noindent Zhiyuan Xie gratefully acknowledges the financial support received from the China Scholarship Council. Gorm Bruun Andresen is funded by the GridScale project supported by the Danish Energy Technology Development and Demonstration Program, Denmark under grant number 64020-2120. Special thanks are extended to Dr. Sleiman Farah, Mr. Ebbe Kyhl Gøtske, and colleagues from the author's focus group for their invaluable assistance with this paper. Appreciation is also due to Miss Fang Fang for her skillful assistance in plotting Figures 1 and 2.

\section*{Declaration of competing interest}
\label{declaration-of-competing-interest}

\noindent The authors declare that they have no known competing financial interests or personal relationships that could have appeared to influence the work reported in this paper.

\section*{Data availability}
\label{data-availability}

\noindent The datasets used as input as well as the data generated by the model have been indicated in the paper.

\section*{Code availability}
\label{code-availability}

\noindent The code about model and plot have been indicated in the paper.

\appendix

\newpage
\section{Key Formulations in PyPSA-Eur-Sec-30 Model}
\label{appendix a}

The PyPSA-Eur-Sec-30-path model jointly optimizes the capacity and dispatch of each asset, including generation, storage, and transmission capacities. This optimization is done under assumptions of perfect foresight, competition, and long-term market equilibrium. Built within the Python for Power System Analysis (PyPSA) framework, the model determines the optimal system configuration by minimizing the total annualized system cost. 

\begin{align}
\min_{\substack{G_{n,s}, E_{n,s}, \\ F_{l}, G_{n,s,t}, F_{l}}} \biggl[ &\sum_{n,s} C_{n,s} \cdot G_{n,s} + \sum_{n,s} \varepsilon_{n,s} \cdot E_{n,s} \nonumber \\
&+ \sum_{l} C_{l} \cdot F_{l} + \sum_{n,s,t} O_{n,s,t} \cdot G_{n,s,t} \biggr] \tag{S1}
\end{align}

\noindent Here, the equation represents the total annualized cost, which is the sum of all costs associated with power and storage capacity, energy storage capacity, bus connectors, and the variable costs of operation across all buses, storage technologies, and time periods. $n$ means the different buses, $s$ means the different technologies, $t$ means the time and resolution is 1h, $l$ include transmission line and converters between the buses implemented in every country (node).

Victoria \textit{et al}.\cite{victoriaearly2020} has identified three potential pathways to decarbonization: linear, a gradual start followed by rapid progression (exponential decay path), and a rapid start followed by a gradual approach (beta distribution path). The latter—beginning swiftly and then slowing down—is considered the best. This paper analysis will directly base on the best path and is as follows:

\begin{equation}
e(t) = e_0 (1 - CDF_{\beta}(t)) \tag{S2}
\end{equation}

\begin{equation}
CDF_{\beta}(t) = \int_{t_0}^{t} PDF_{\beta}(t)dt \tag{S3}
\end{equation}

\begin{equation}
PDF_{\beta}(t) = \frac{\Gamma(2\beta)}{2\Gamma(\beta)^2} (t - t_0)^{\beta-1}(t_f - t)^{\beta-1} \tag{S4}
\end{equation}

\noindent In 2020, carbon emissions from electricity generation and the heating supply in Europe's residential and service sectors were 1.56 GtCO$_2$ , constituting $43.5\%$ of total European emissions. For the future, the carbon budget for these sectors in Europe is determined to be 21 GtCO$_2$ . This budget is estimated in the context of a global carbon budget of 800 GtCO$_2$, which is the threshold to prevent temperature increases above $1.75°C$ relative to the preindustrial era with a probability exceeding $66\%$ \cite{IPCC2018}. The integral from the base year $t_0$ extending into the future of $e(t)$ is set to equal this carbon budget B, $\int_{t_0}^{\infty} e(t) \, dt = B$. That will mean we have a CO$_2$ emission cap for each time step. 

This optimization process also should subject to various constraints, and among these constraints, the match of the supply and demand is the most important.

\begin{equation}
\sum_{s} G_{n,s,t} + \sum_{l} a_{n,l,t} \cdot f_{l,t} = d_{n,t} \quad \leftrightarrow \quad \lambda_{n,t} \quad \forall n, t \tag{S5}
\end{equation}

\noindent In the energy system model, $\lambda_{n,t}$ represents the Karush-Kuhn-Tucker (KKT) multiplier. When equilibrium is achieved in the electricity bus, this multiplier reflects the locational marginal price (LMP), which is the price of electricity at that specific node. 

\section{Supplementary Figures}
\label{appendix b}

The following pages present a series of Supplementary Figures, numbered S1 through S25. Each figure is provided on a separate page for clarity and detailed examination.

\setcounter{figure}{0}
\renewcommand{\thefigure}{S\arabic{figure}}

\begin{figure*}[t]
	\centering 
	\includegraphics[width=0.9\textwidth, angle=0]{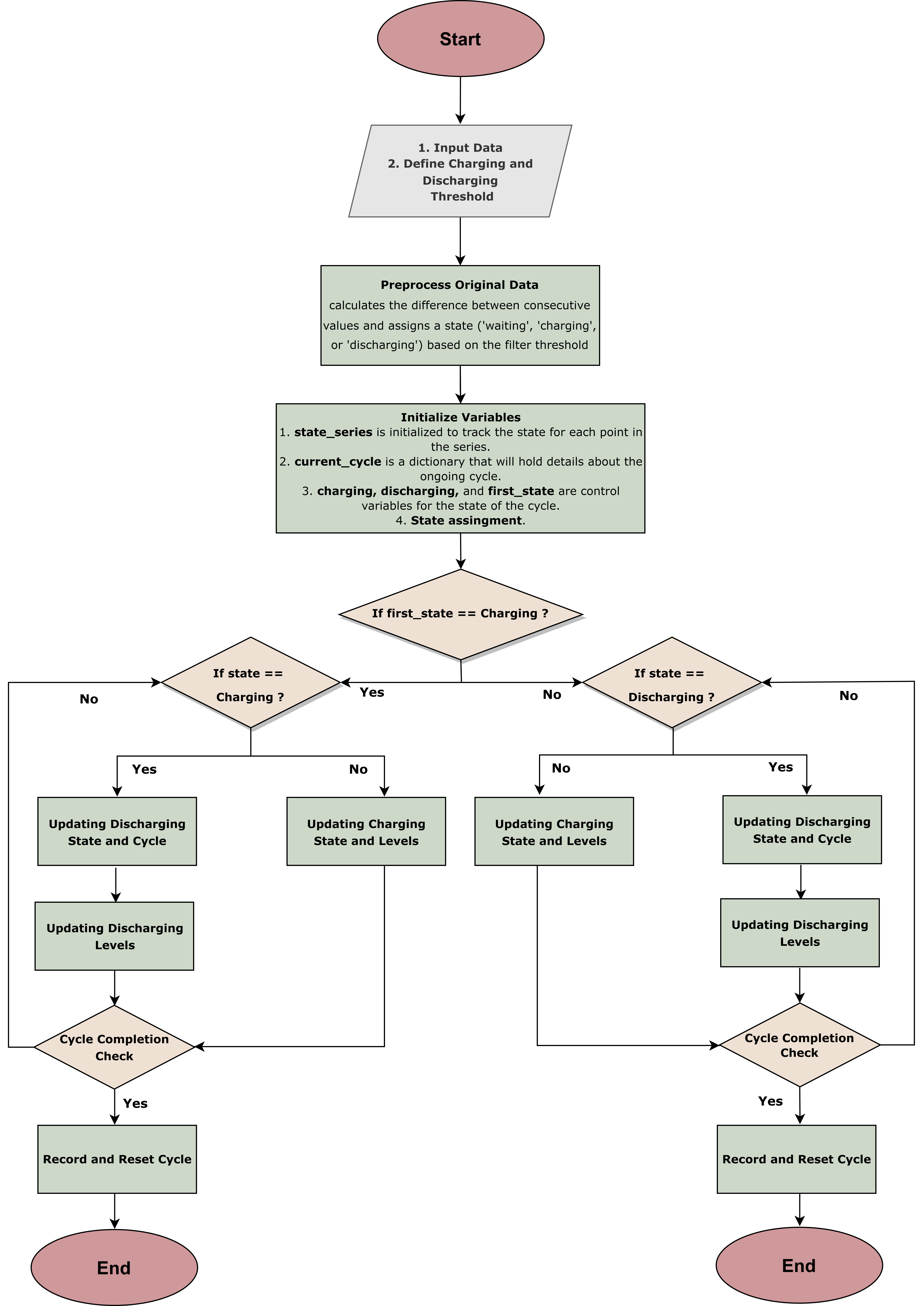}	
	\caption{Operation Logic of Cycle Capture Method} 
	\label{fig_S1}
\end{figure*}

\begin{figure*}[t]
	\centering 
	\includegraphics[width=0.9\textwidth, angle=0]{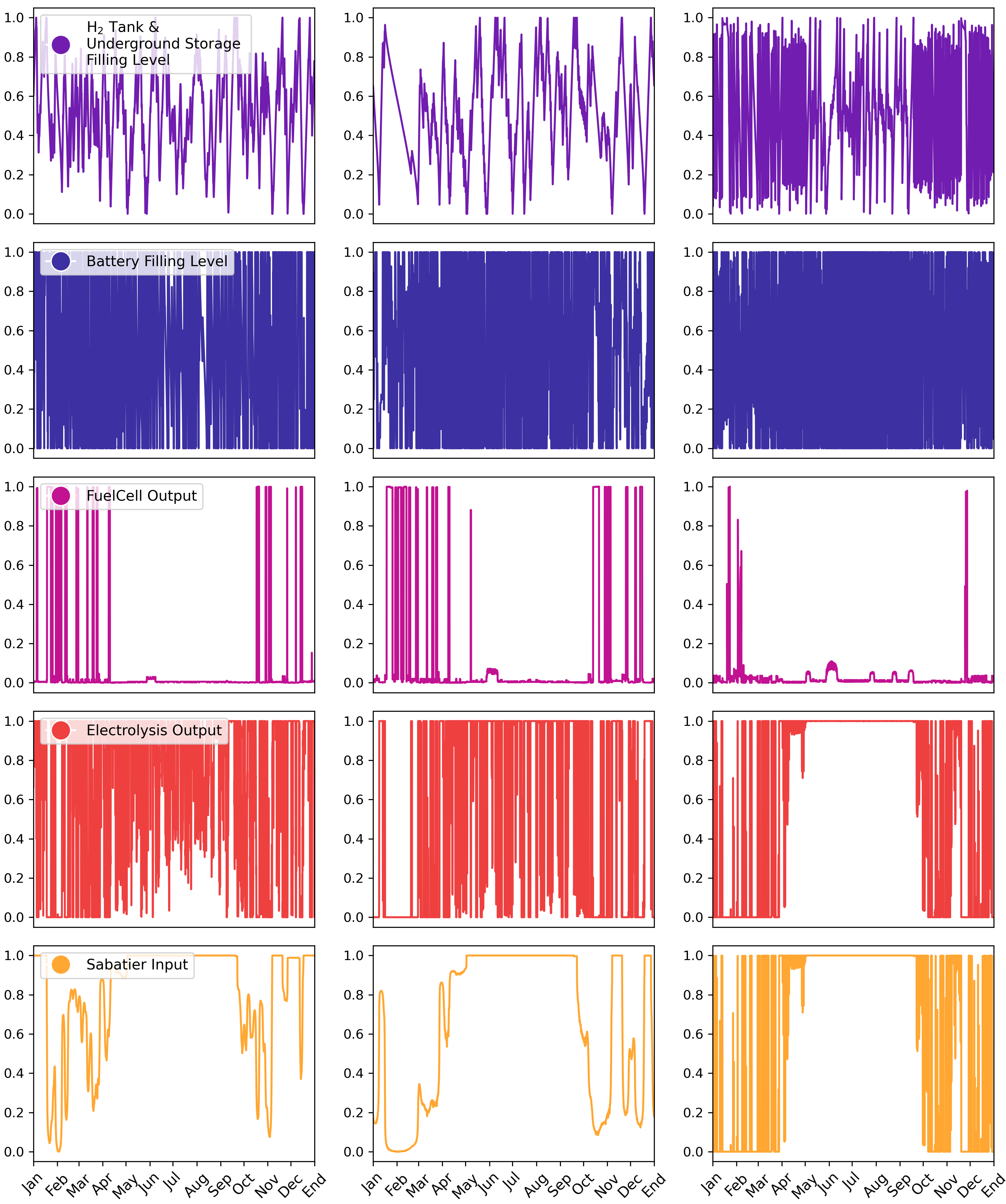}	
	\caption{Time Series Analysis of Denmark, Germany and Italy Hydrogen Sector and Battery Storage for 2050} 
	\label{fig_S2}
\end{figure*}

\begin{figure*}[t]
	\centering 
	\includegraphics[width=\textwidth, angle=0]{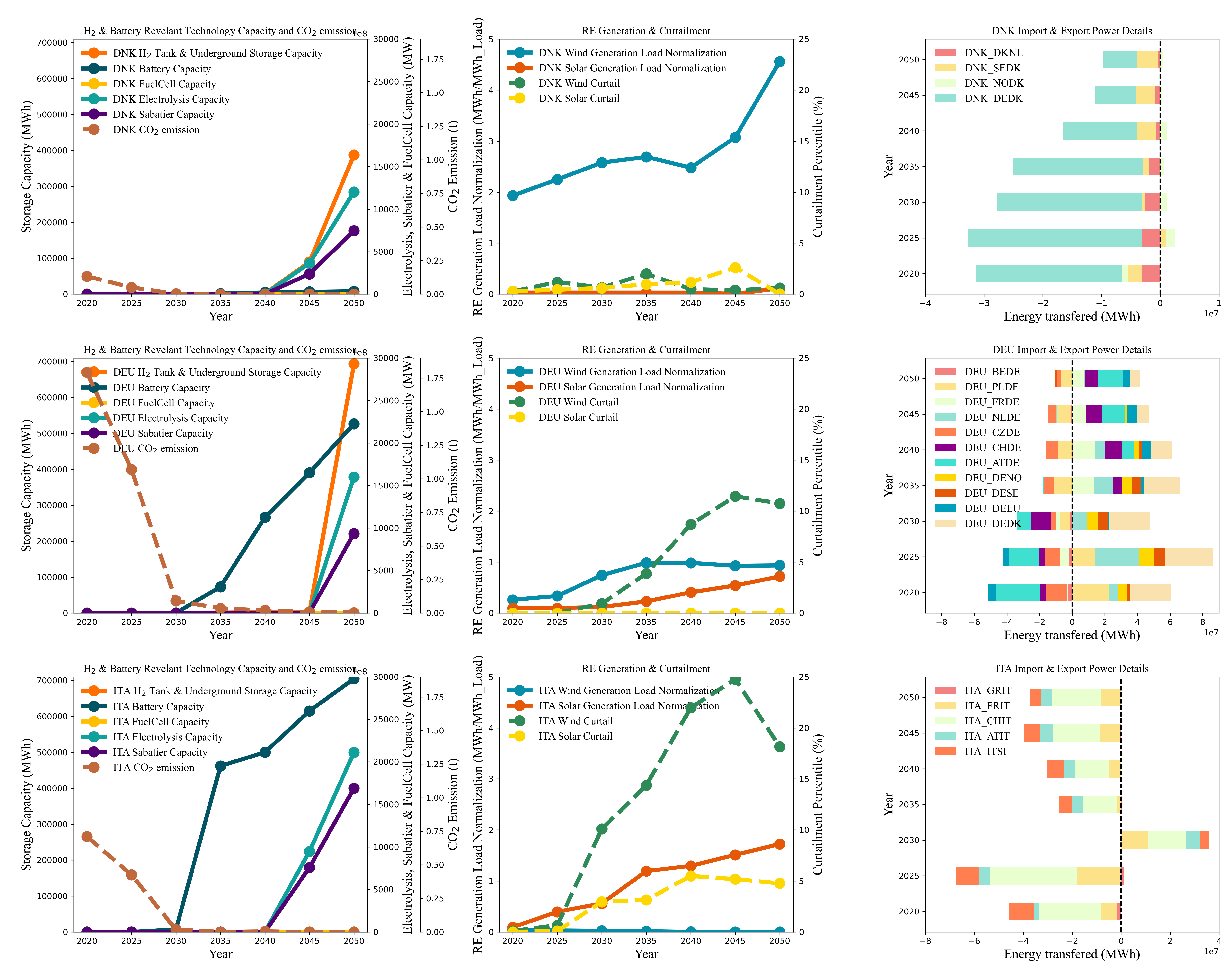}	
	\caption{Evolution of Renewable Energy Landscapes in Denmark, Germany, and Italy: Capacity, Generation, Penetration, Cross-country Flow of Electricity, and CO$_2$ Emissions (2020-2050)} 
	\label{fig_S3}
\end{figure*}

\begin{figure*}[t]
	\centering 
	\includegraphics[width=1\textwidth, angle=0]{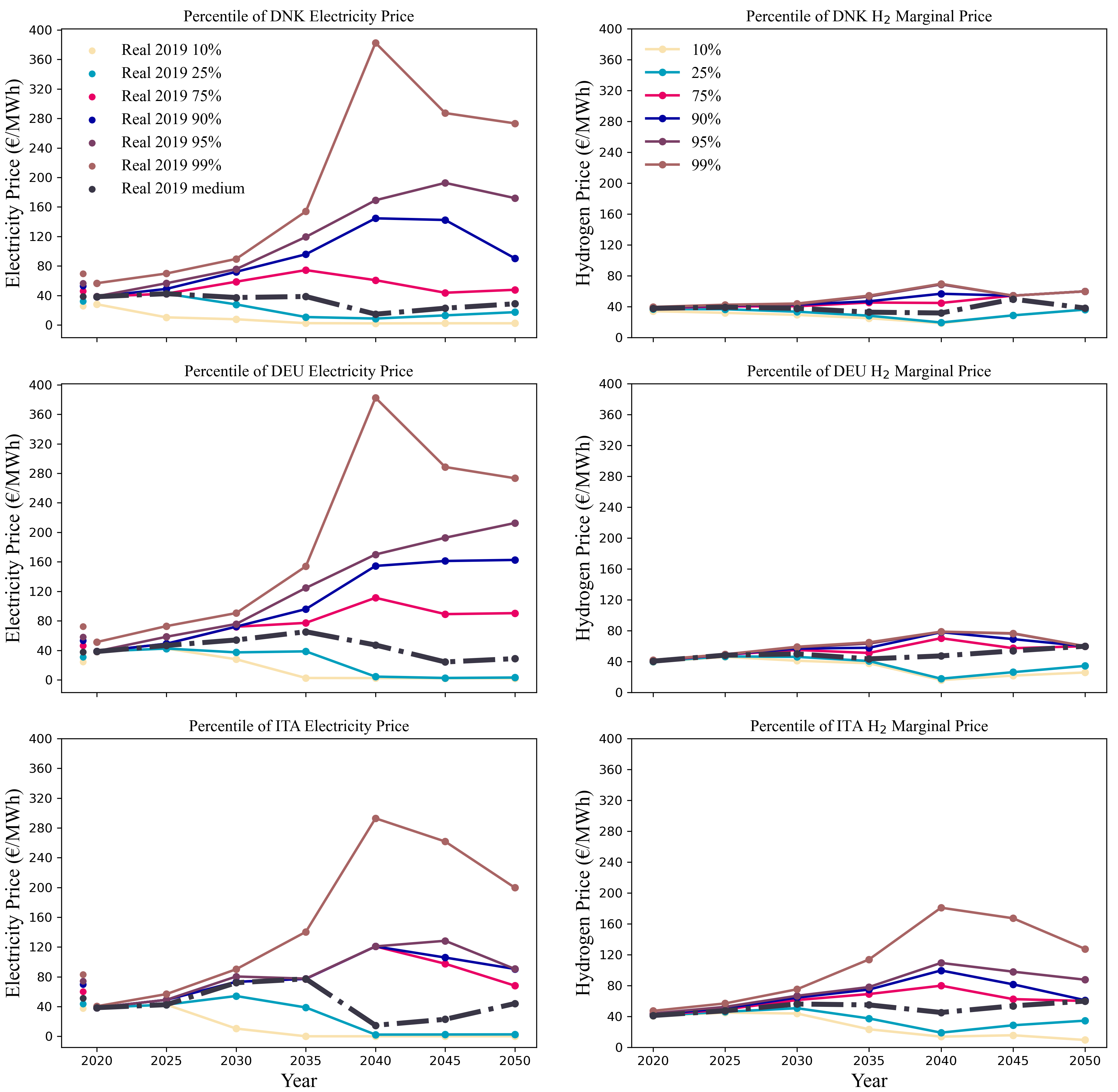}	
	\caption{Historical and Projected Electricity Price Percentiles in Denmark, Germany, and Italy: A 2050 Outlook} 
	\label{fig_S4}
\end{figure*}

\begin{figure*}[t]
	\centering 
	\includegraphics[width=0.9\textwidth, angle=0]{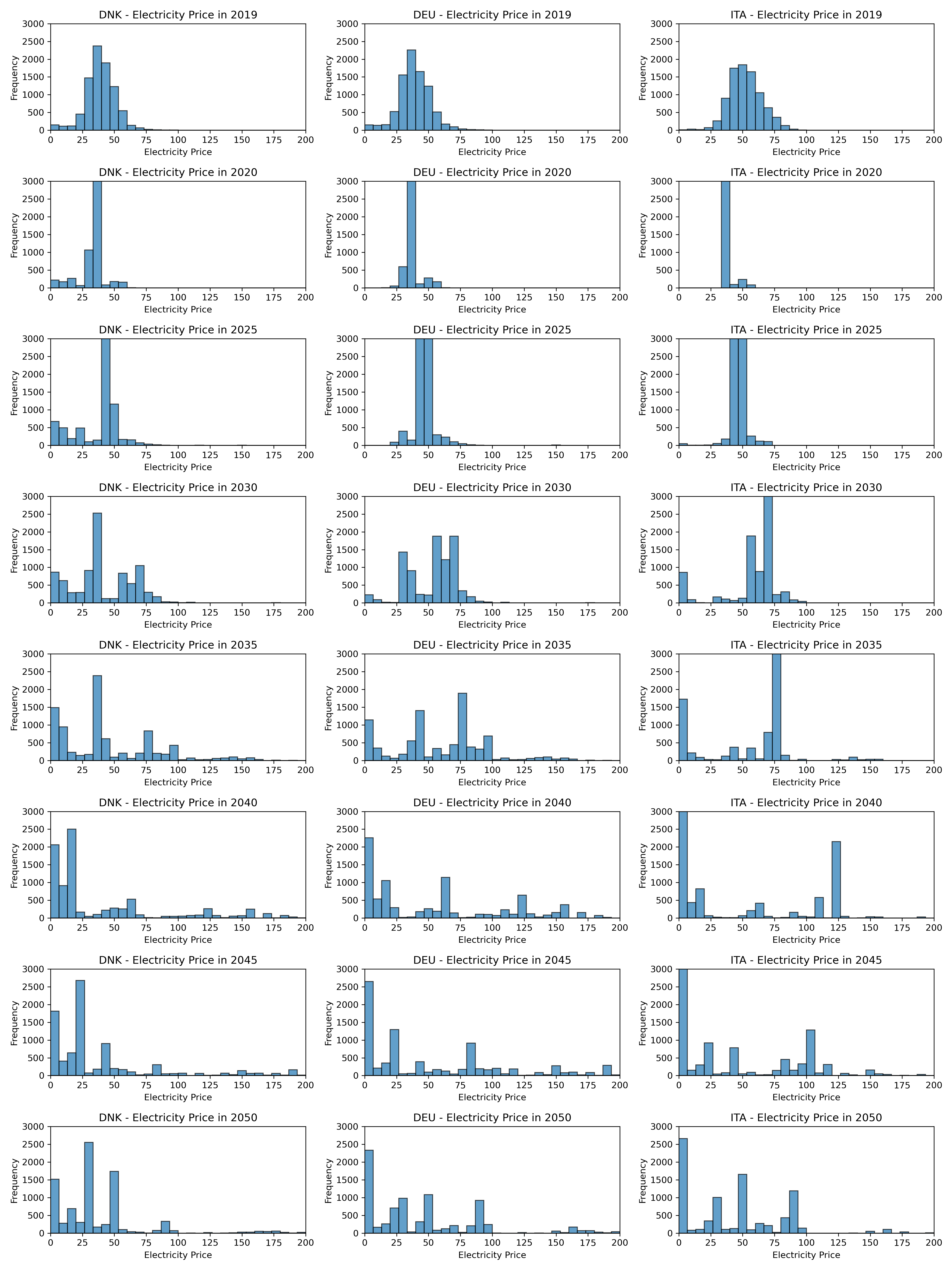}	
	\caption{Historical and Projected Electricity Price Distribution in Denmark, Germany, and Italy: A 2050 Outlook} 
	\label{fig_S5}
\end{figure*}

\begin{figure*}[t]
	\centering 
	\includegraphics[width=0.9\textwidth, angle=0]{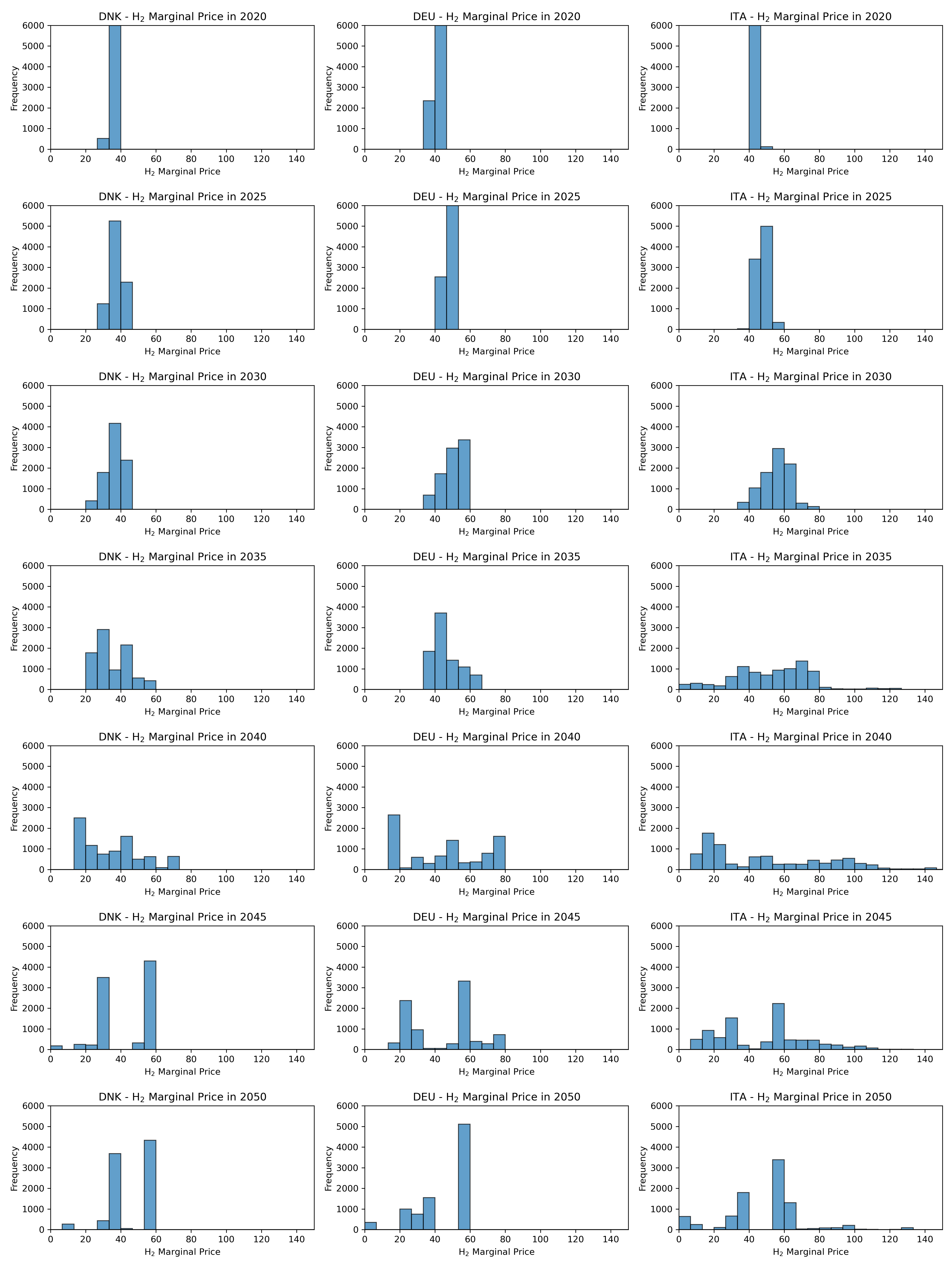}	
	\caption{Historical and Projected H$_2$ Marginal Price Distribution in Denmark, Germany, and Italy: A 2050 Outlook} 
	\label{fig_S6}
\end{figure*}

\begin{figure*}[t]
	\centering 
	\includegraphics[width=\textwidth, angle=0]{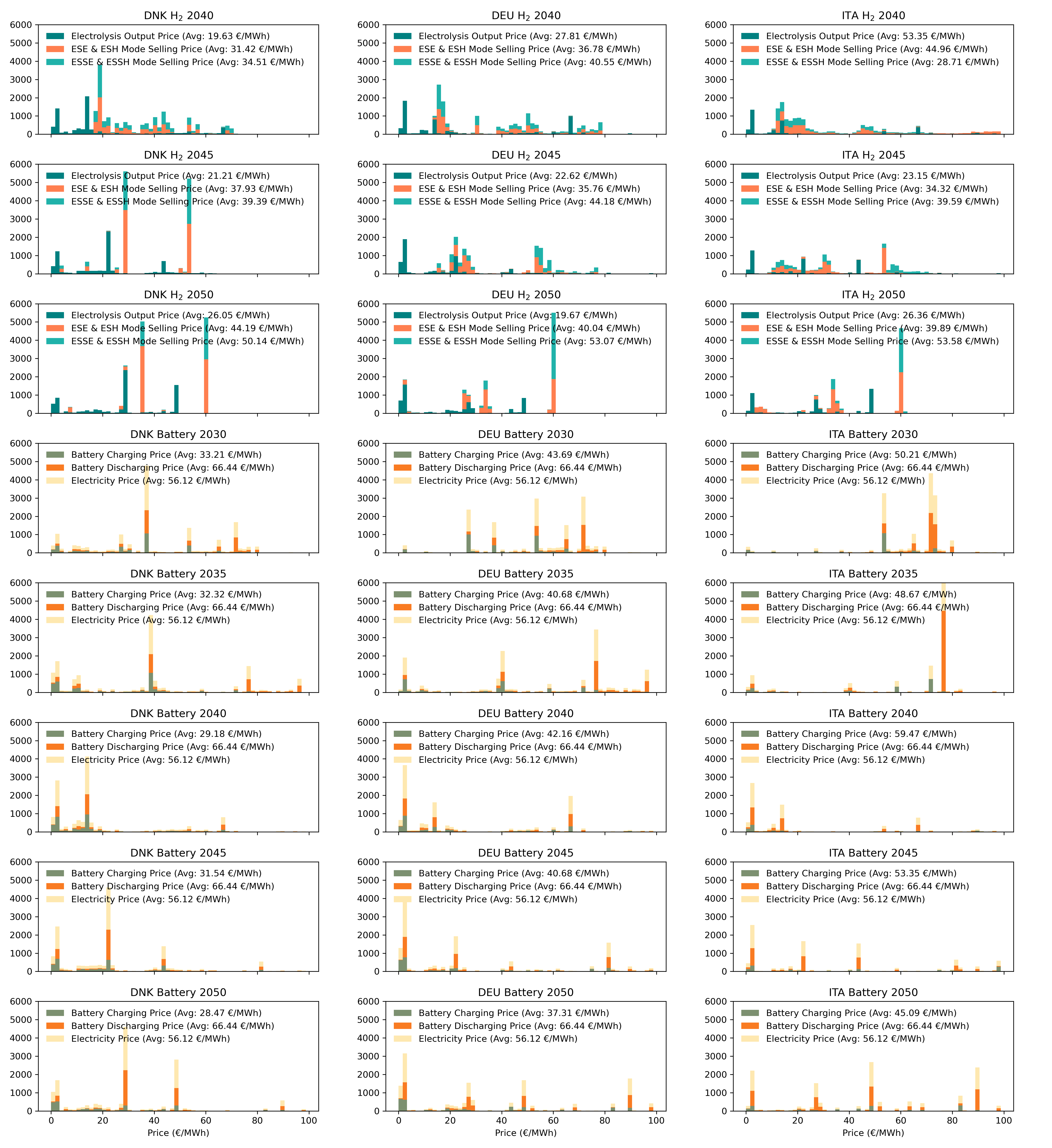}	
	\caption{Comparative Price Distribution for Battery Storage Buying and Selling and Electricity (2030-2050) with Detailed Hydrogen Buying and Selling Analysis (2040-2050) in Denmark, Germany, and Italy} 
	\label{fig_S7}
\end{figure*}

\begin{figure*}[t]
	\centering 
	\includegraphics[width=.95\textwidth, angle=0]{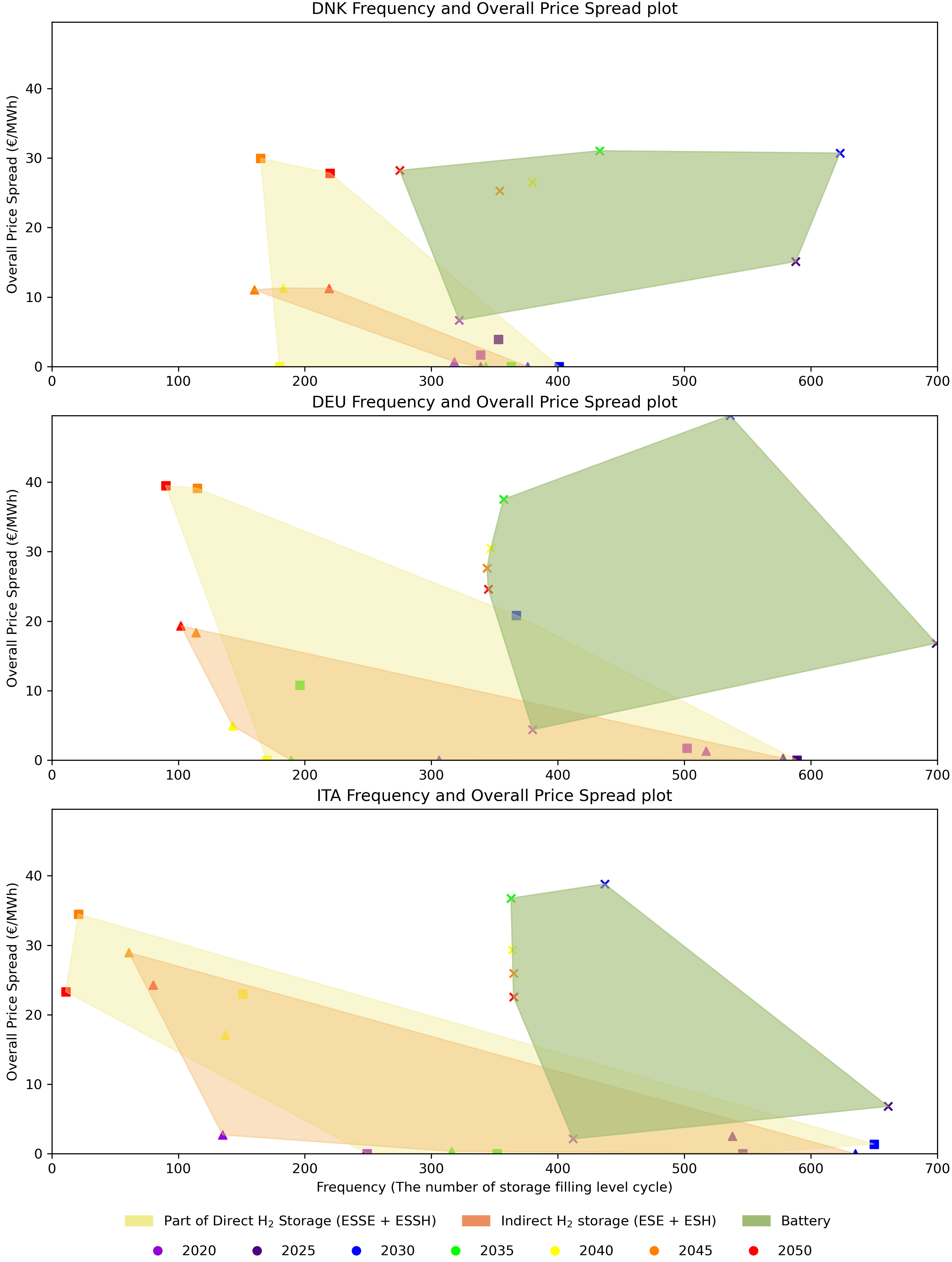}	
	\caption{Comparative Performance of H$_2$ and Battery Systems in Denmark and Italy} 
	\label{fig_S8}
\end{figure*}

\begin{figure*}[t]
	\centering 
	\includegraphics[width=1.15\textwidth, angle=90]{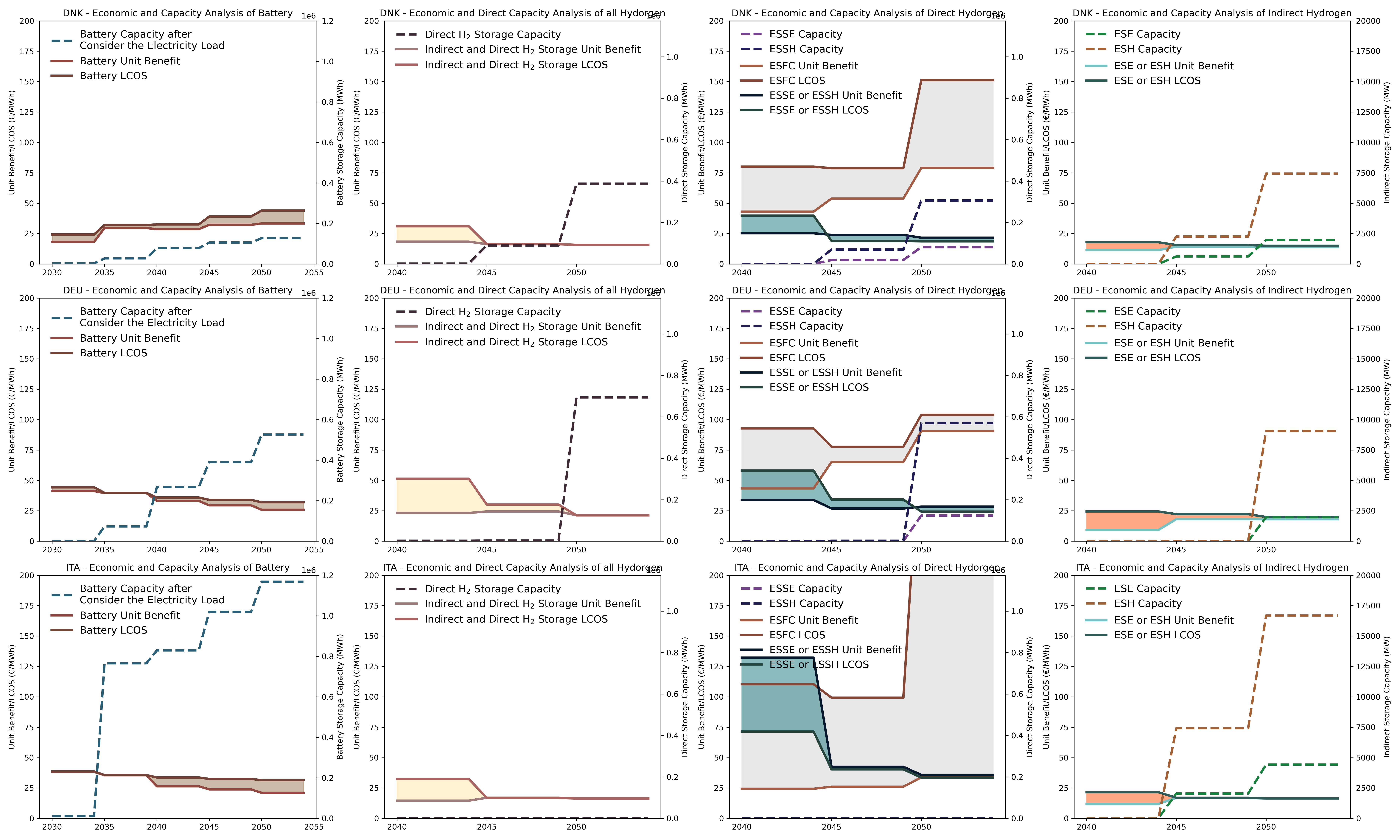}	
	\caption{Projected Trends in Accumulated LCOS, Unit Benefits, and Capacities for Hydrogen and Battery Storage in Denmark, Germany, and Italy in the Late Energy Transition Stage} 
	\label{fig_S9}
\end{figure*}

\begin{figure*}[t]
	\centering 
	\includegraphics[width=1.15\textwidth, angle=90]{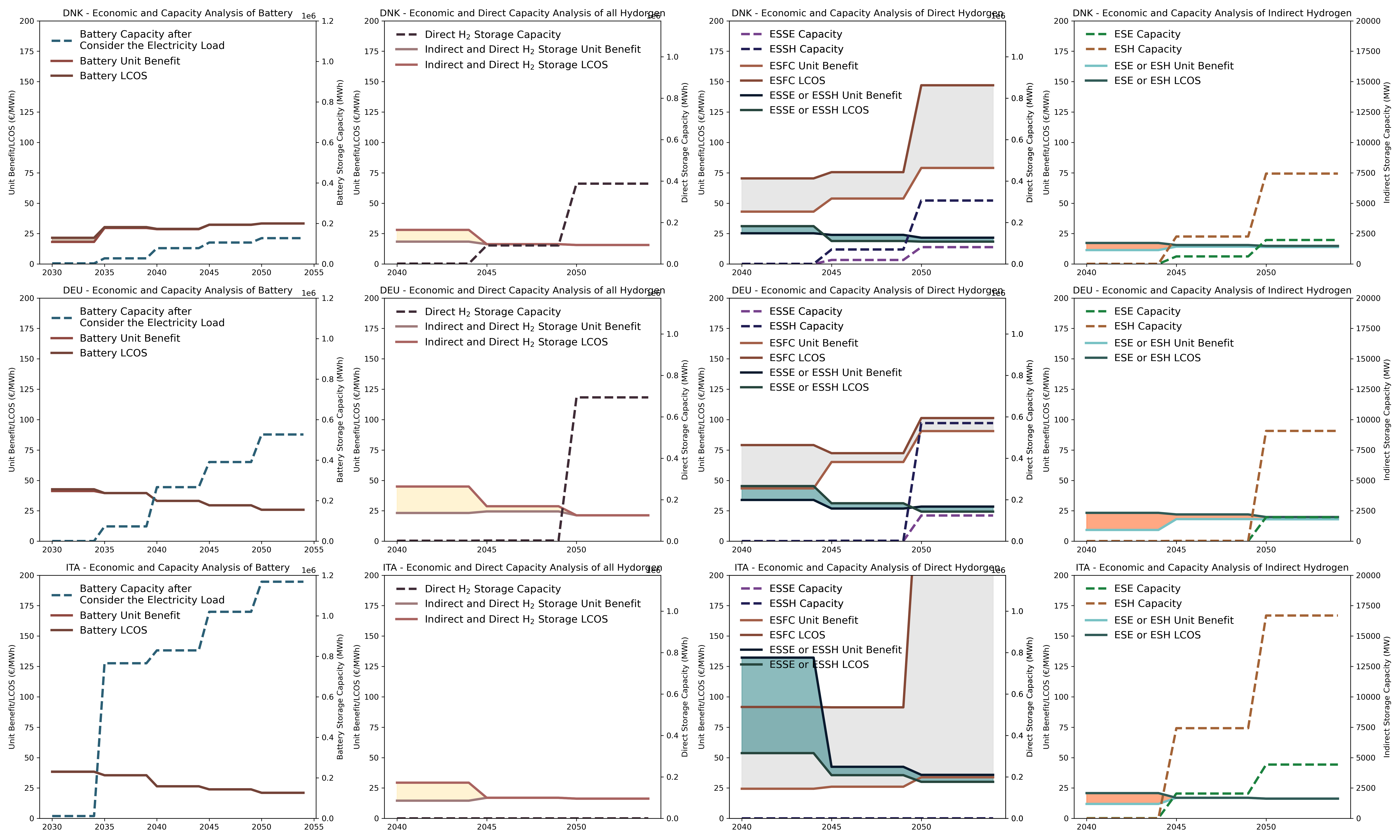}	
	\caption{Comparative Analysis of Non-Accumulated LCOS for Hydrogen and Battery Storage in Denmark, Germany, and Italy in the Late Energy Transition Stage} 
	\label{fig_S10}
\end{figure*}

\begin{figure*}[t]
	\centering 
	\includegraphics[width=.95\textwidth, angle=0]{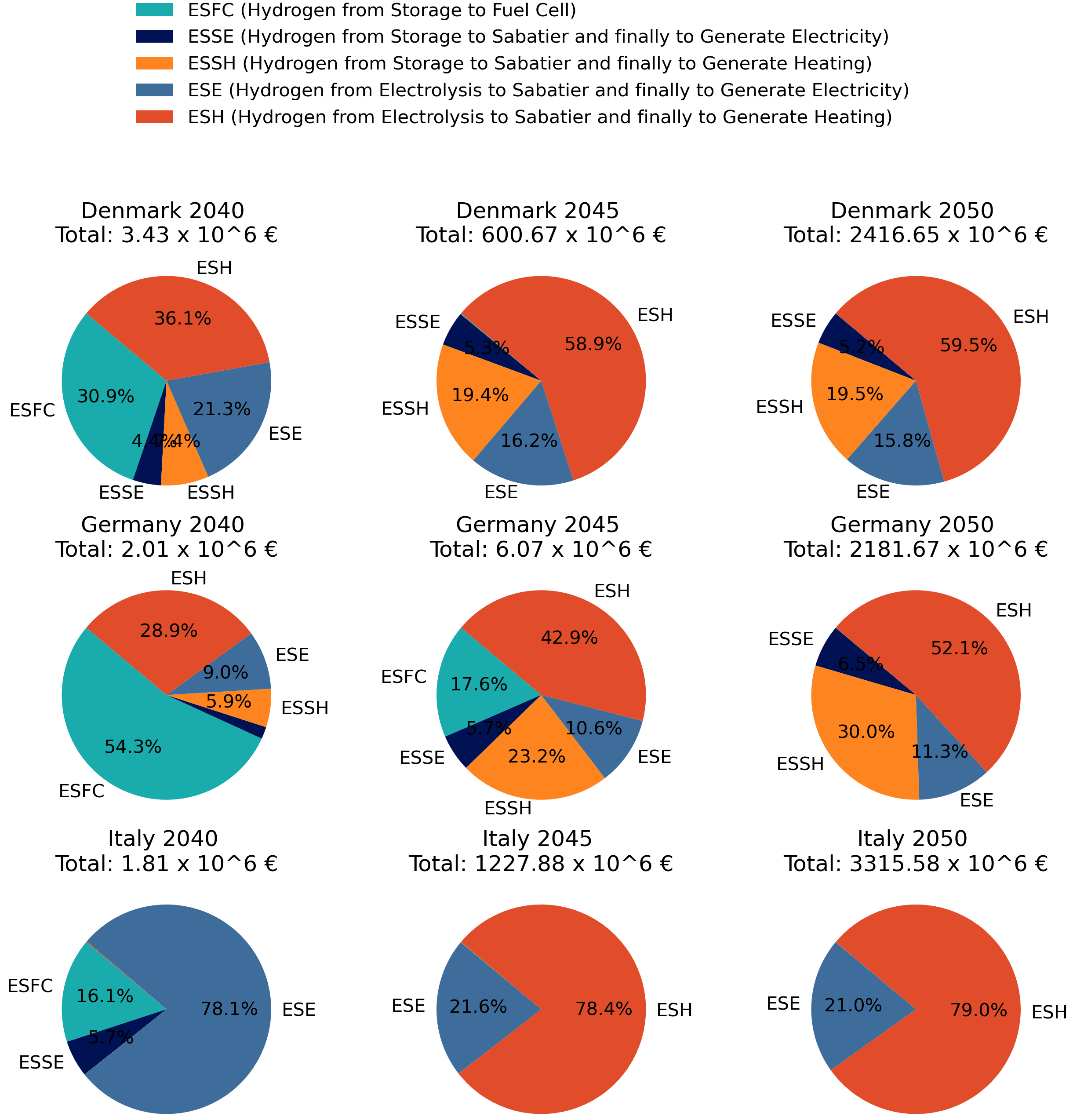}	
	\caption{Revenue Distribution of Denmark, Germany and Italy H$_2$ Storage Working Modes in 2040, 2045, and 2050} 
	\label{fig_S11}
\end{figure*}

\begin{figure*}[t]
	\centering 
	\includegraphics[width=1.2\textwidth, angle=90]{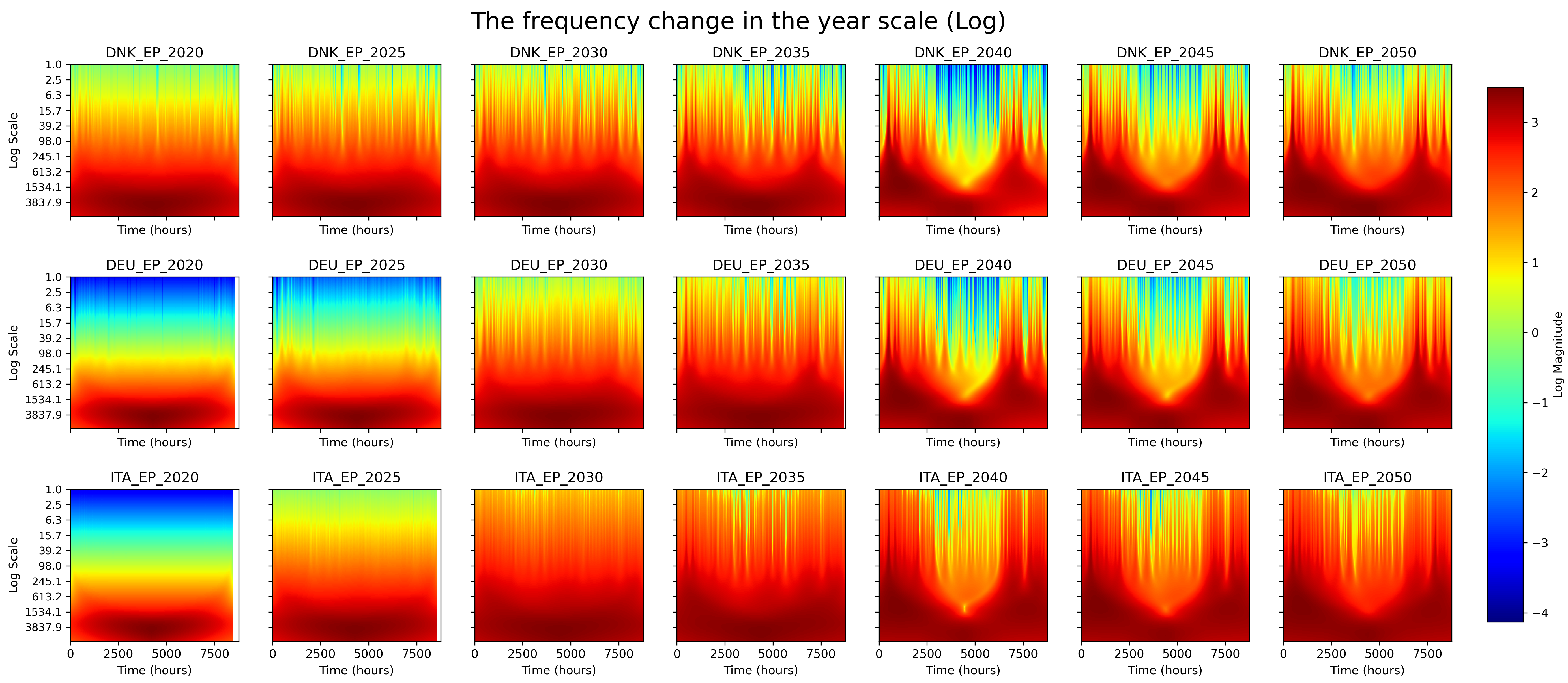}	
	\caption{Continuous Wavelet Transform (CWT) Analysis of Electricity Prices in Denmark, Germany, and Italy} 
	\label{fig_S12}
\end{figure*}

\begin{figure*}[t]
	\centering 
	\includegraphics[width=0.97\textwidth, angle=0]{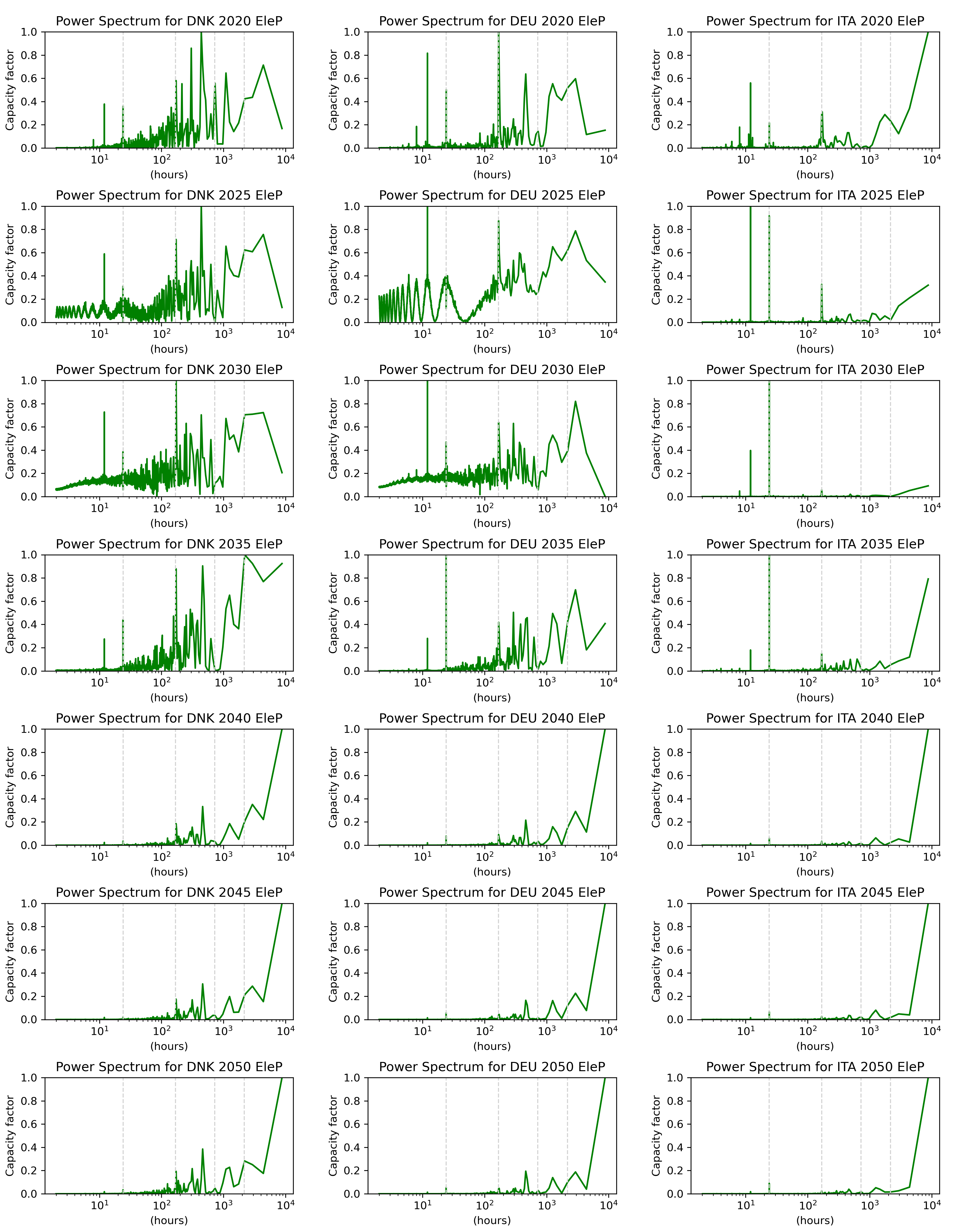}	
	\caption{Fast Fourier Transform (FFT) Analysis of Electricity Prices in Denmark, Germany, and Italy} 
	\label{fig_S13}
\end{figure*}

\begin{figure*}[t]
	\centering 
	\includegraphics[width=0.9\textwidth, angle=0]{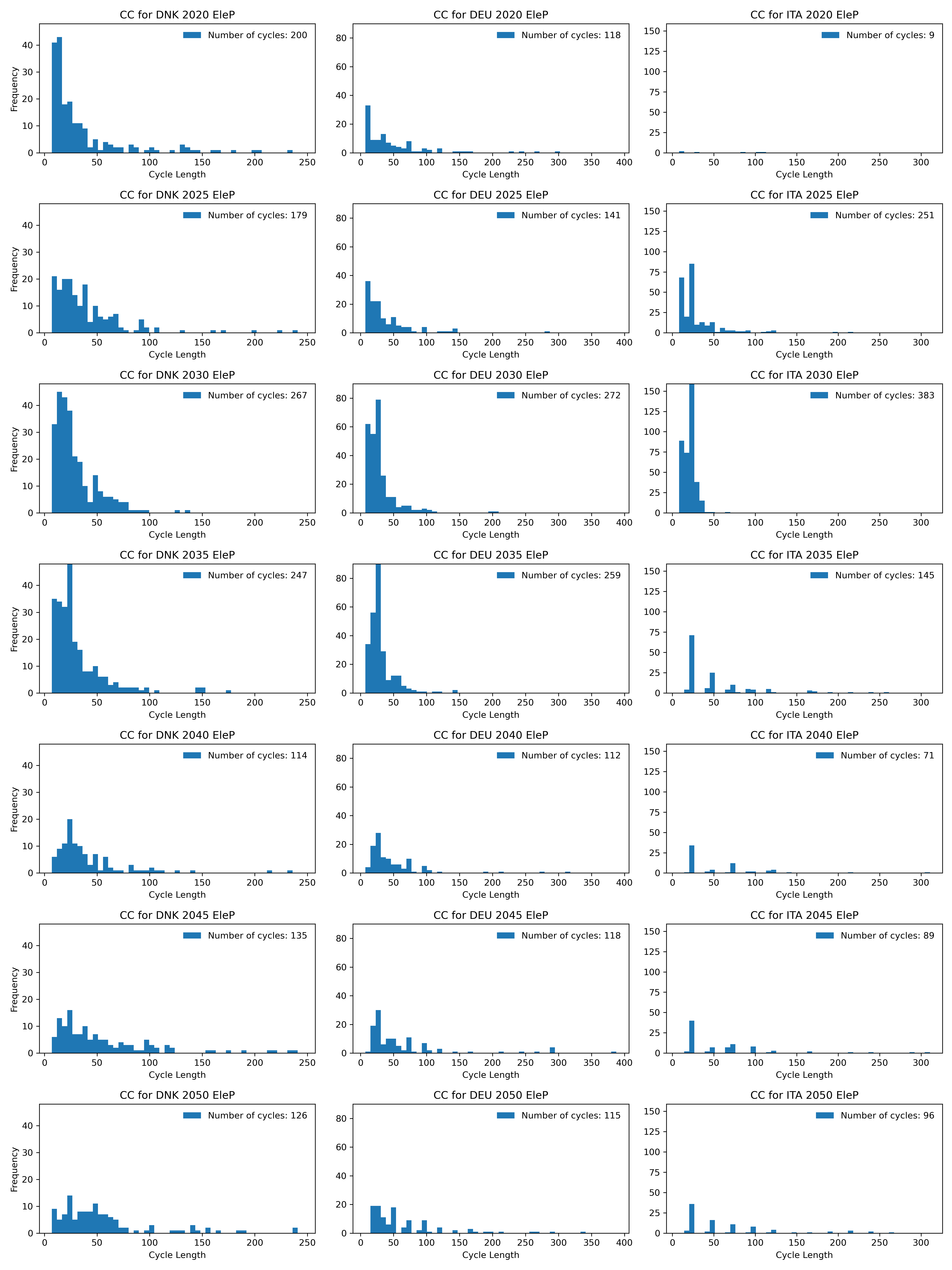}	
	\caption{Cycle Capture (CC) Analysis of Electricity Prices in Denmark, Germany, and Italy} 
	\label{fig_S14}
\end{figure*}

\begin{figure*}[t]
	\centering 
	\includegraphics[width=1.02\textwidth, angle=0]{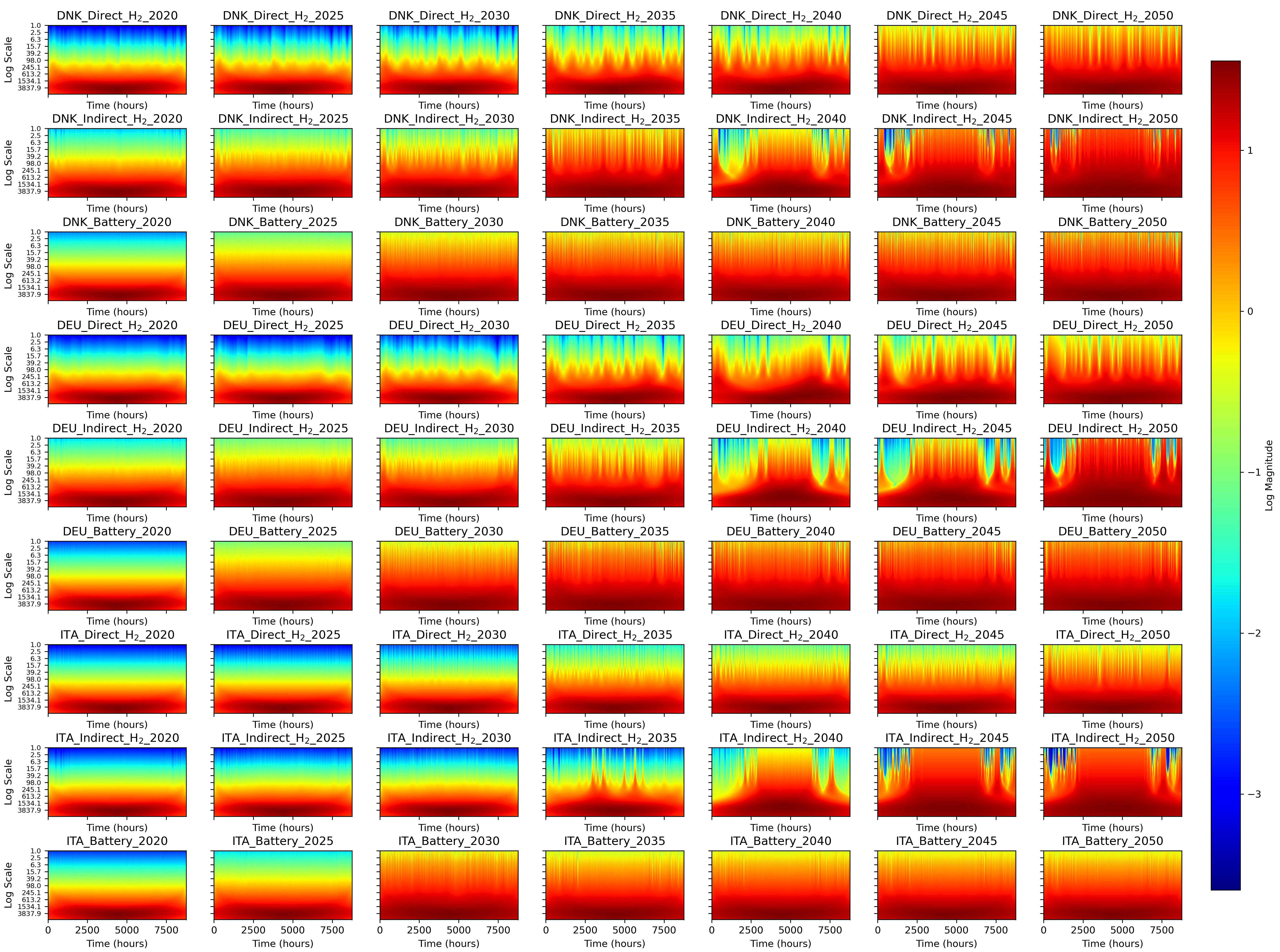}	
	\caption{Continuous Wavelet Transform (CWT) Analysis of Storage Filling Level in Denmark, Germany, and Italy} 
	\label{fig_S15}
\end{figure*}

\begin{figure*}[t]
	\centering 
	\includegraphics[width=\textwidth, angle=0]{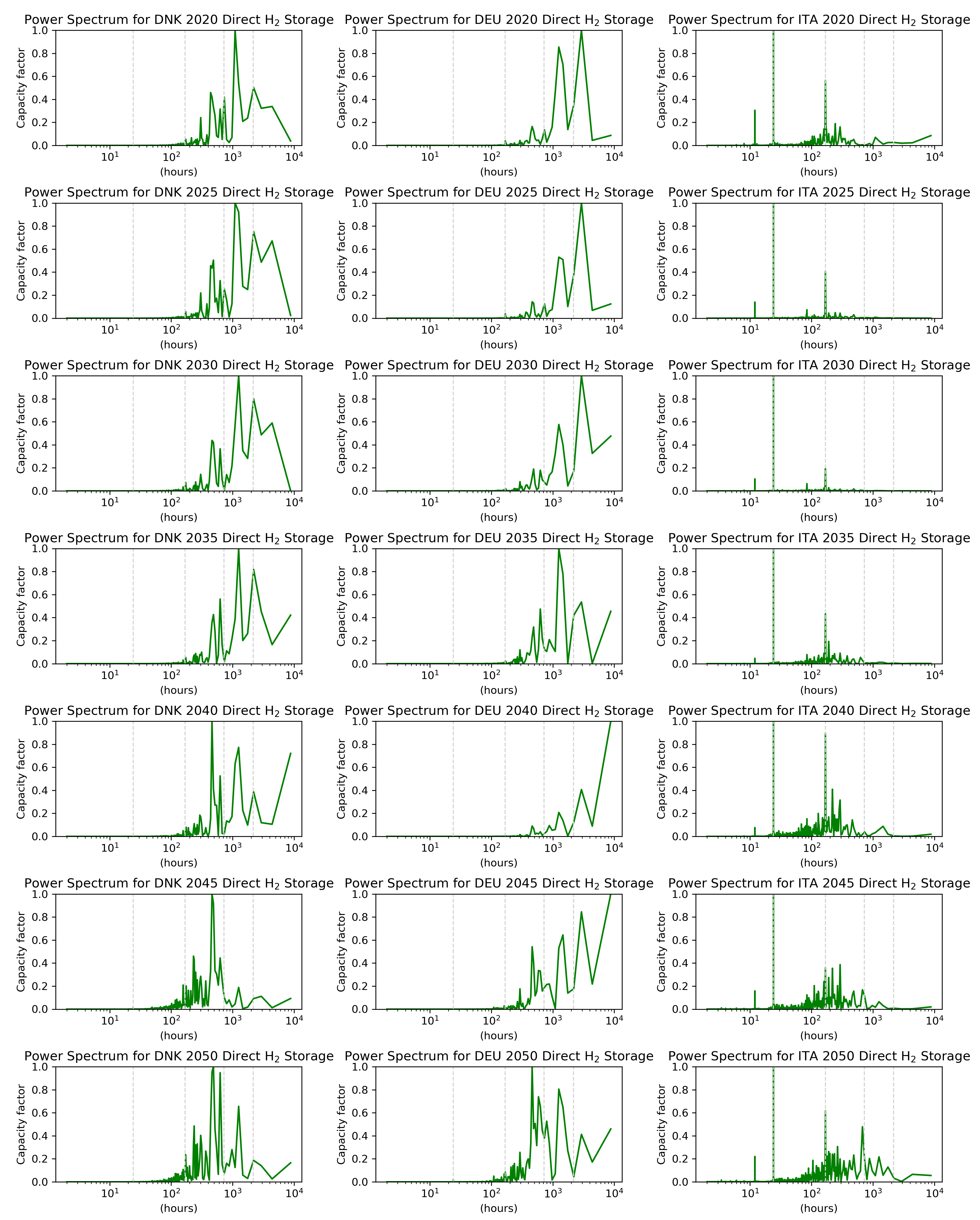}	
	\caption{Fast Fourier Transform (FFT) Analysis of Direct Hydrogen Storage Filling Levels in Denmark, Germany, and Italy} 
	\label{fig_S16}
\end{figure*}

\begin{figure*}[t]
	\centering 
	\includegraphics[width=0.9\textwidth, angle=0]{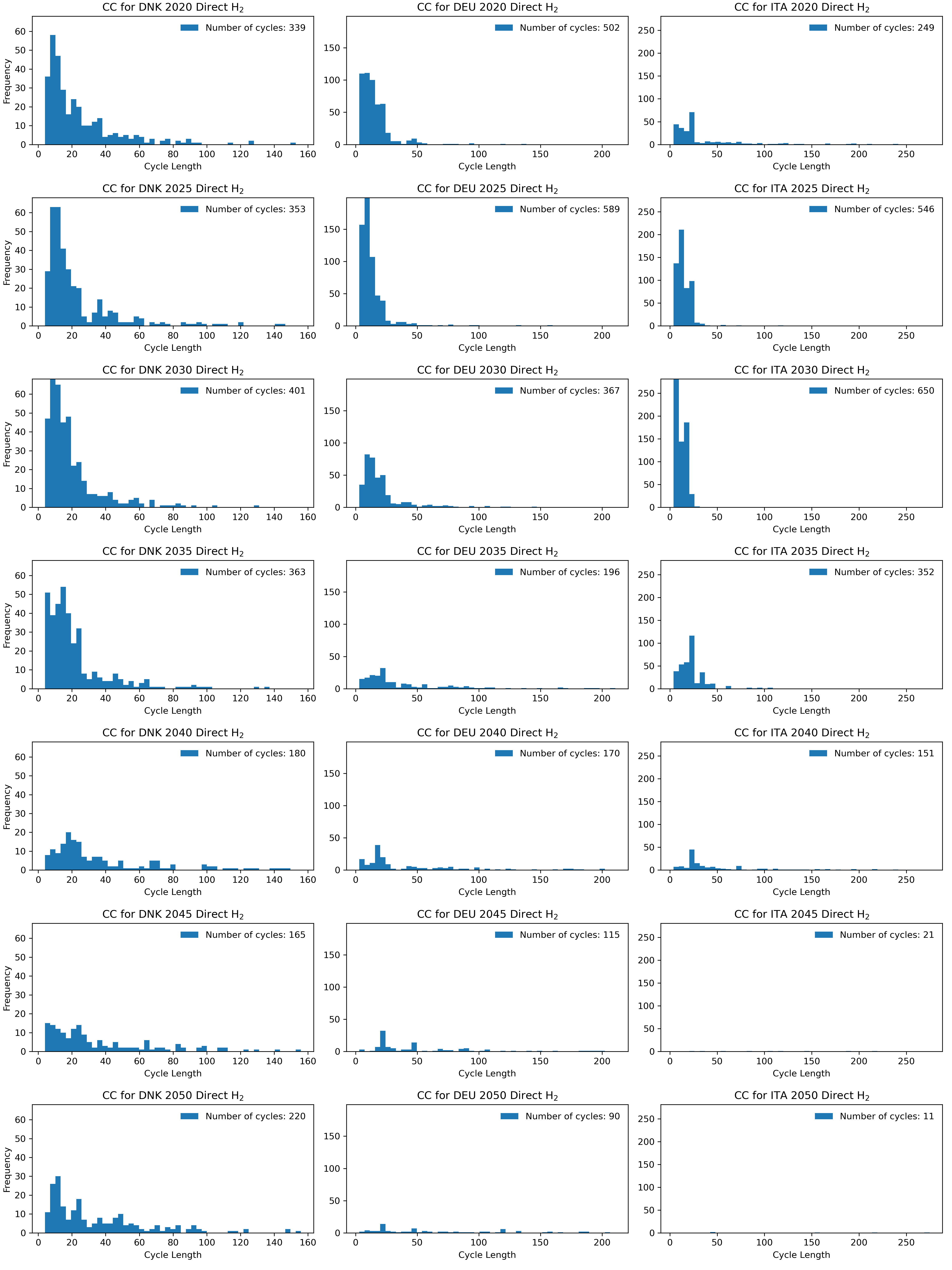}	
	\caption{Cycle Capture (CC) Analysis of Direct Hydrogen Storage Filling Levels in Denmark, Germany, and Italy} 
	\label{fig_S17}
\end{figure*}

\begin{figure*}[t]
	\centering 
	\includegraphics[width=.97\textwidth, angle=0]{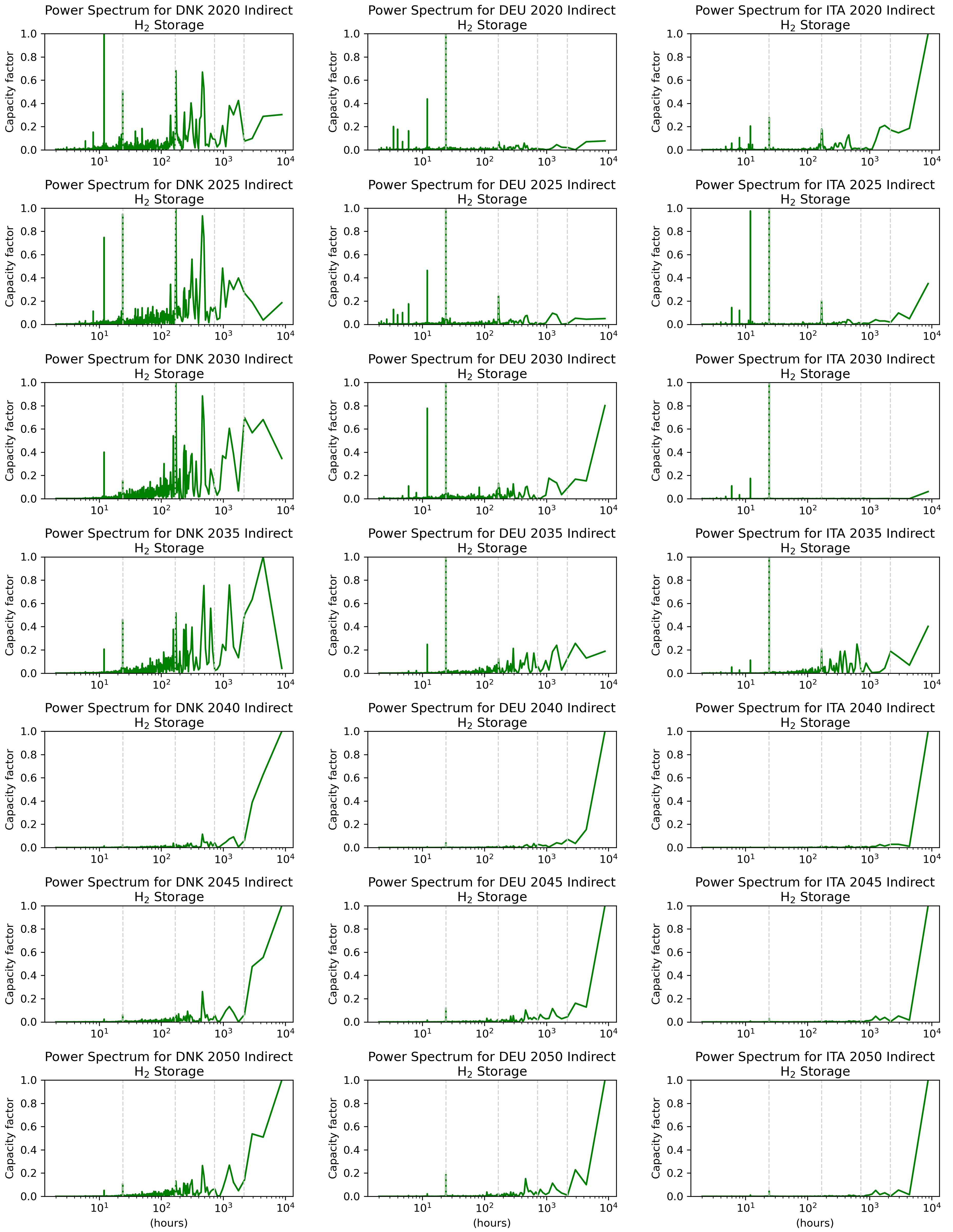}	
	\caption{Fast Fourier Transform (FFT) Analysis of Indirect Hydrogen Storage Filling Levels in Denmark, Germany, and Italy} 
	\label{fig_S18}
\end{figure*}

\begin{figure*}[t]
	\centering 
	\includegraphics[width=0.9\textwidth, angle=0]{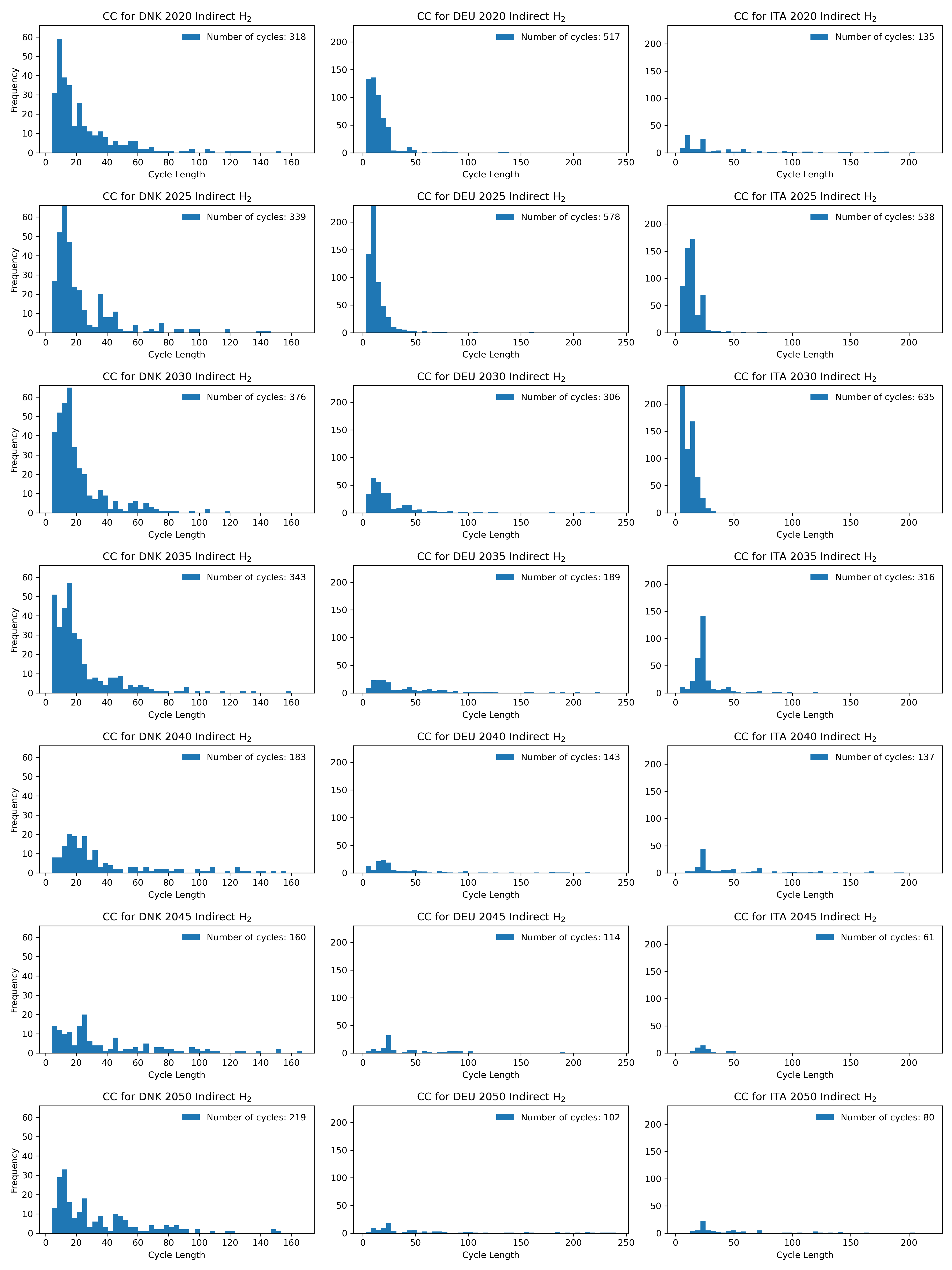}	
	\caption{Cycle Capture (CC) Analysis of Indirect Hydrogen Storage Filling Levels in Denmark, Germany, and Italy} 
	\label{fig_S19}
\end{figure*}

\begin{figure*}[t]
	\centering 
	\includegraphics[width=0.9\textwidth, angle=0]{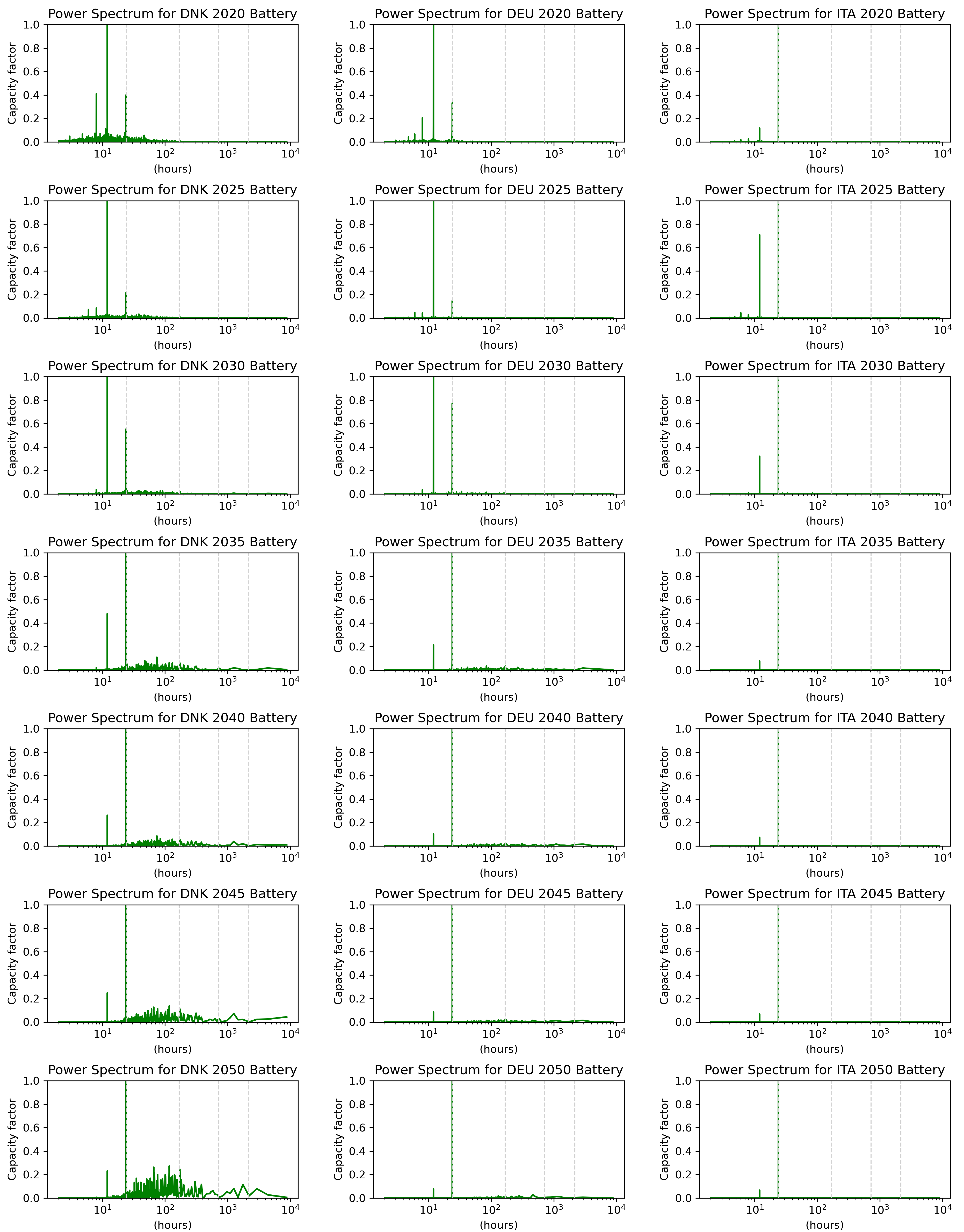}	
	\caption{Fast Fourier Transform (FFT) Analysis of Battery Filling Levels in Denmark, Germany, and Italy} 
	\label{fig_S20}
\end{figure*}

\begin{figure*}[t]
	\centering 
	\includegraphics[width=.95\textwidth, angle=0]{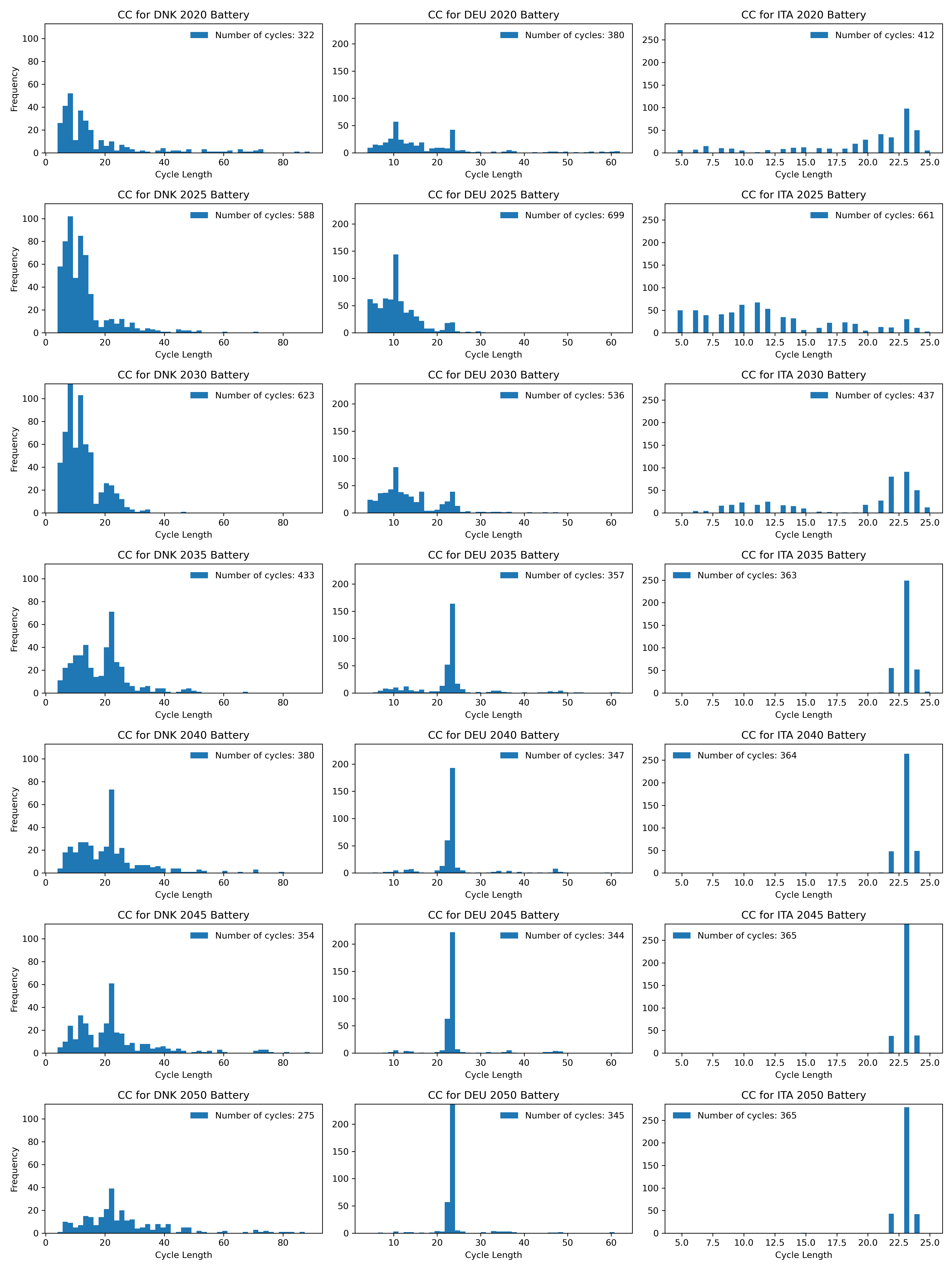}	
	\caption{Cycle Capture (CC) Analysis of Battery Filling Levels in Denmark, Germany, and Italy} 
	\label{fig_S21}
\end{figure*}

\begin{figure*}[t]
	\centering 
	\includegraphics[width=1.2\textwidth, angle=90]{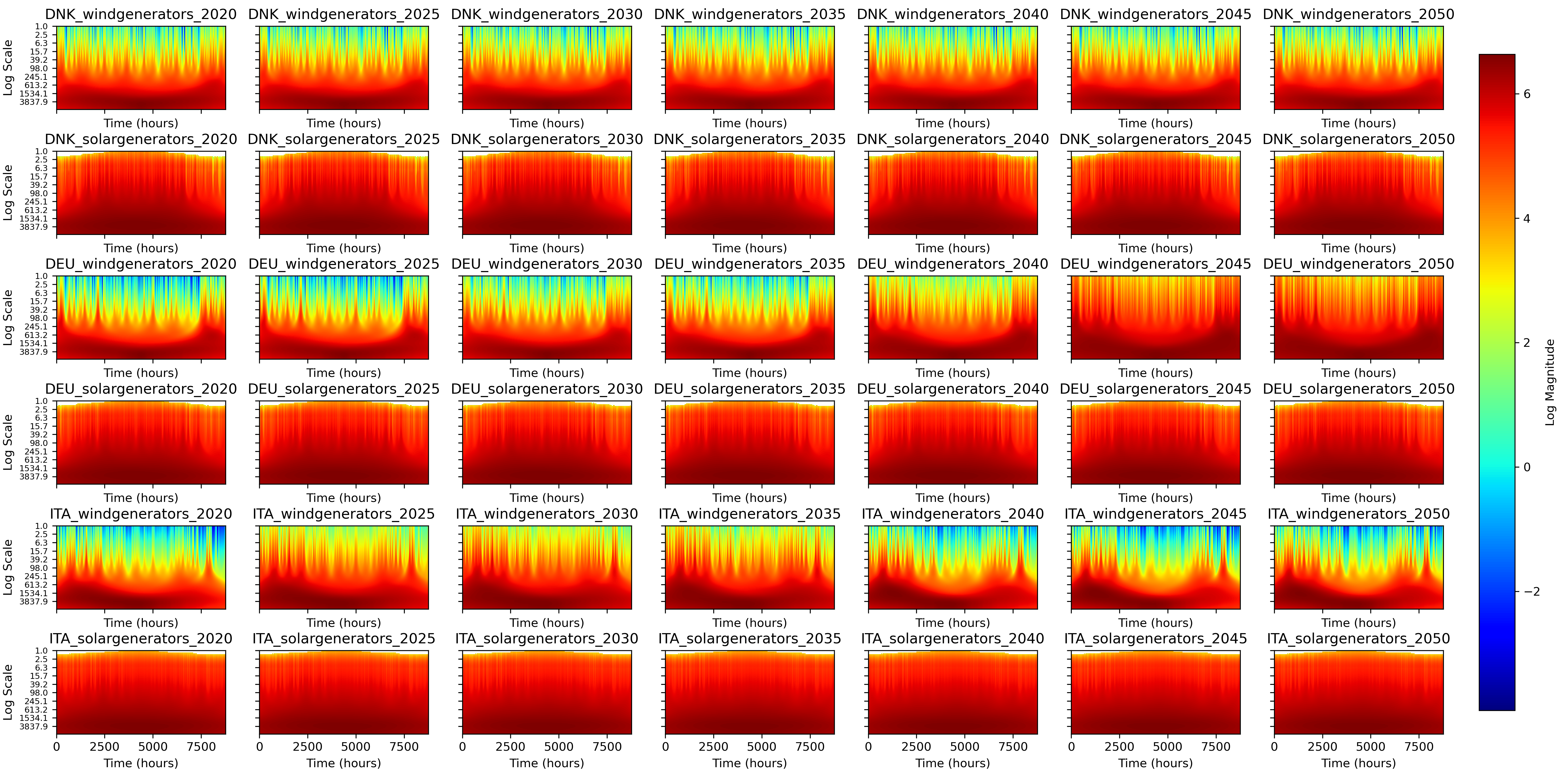}	
	\caption{Continuous Wavelet Transform (CWT) Analysis of Renewable Energy Generation in Denmark, Germany, and Italy} 
	\label{fig_S22}
\end{figure*}

\begin{figure*}[t]
	\centering 
	\includegraphics[width=.97\textwidth, angle=0]{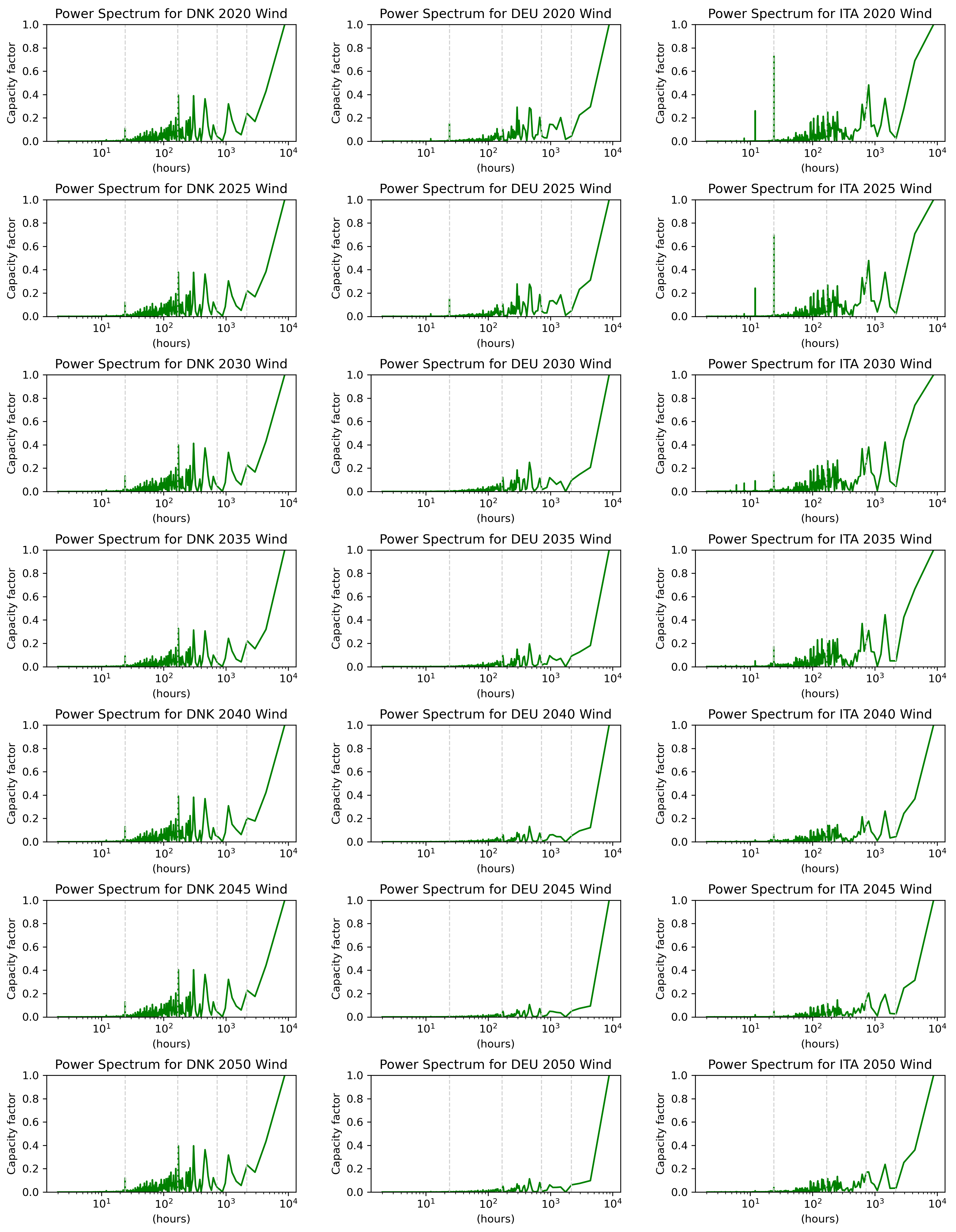}	
	\caption{Fast Fourier Transform (FFT) Analysis of Wind Generations in Denmark, Germany, and Italy} 
	\label{fig_S23}
\end{figure*}

\begin{figure*}[t]
	\centering 
	\includegraphics[width=0.9\textwidth, angle=0]{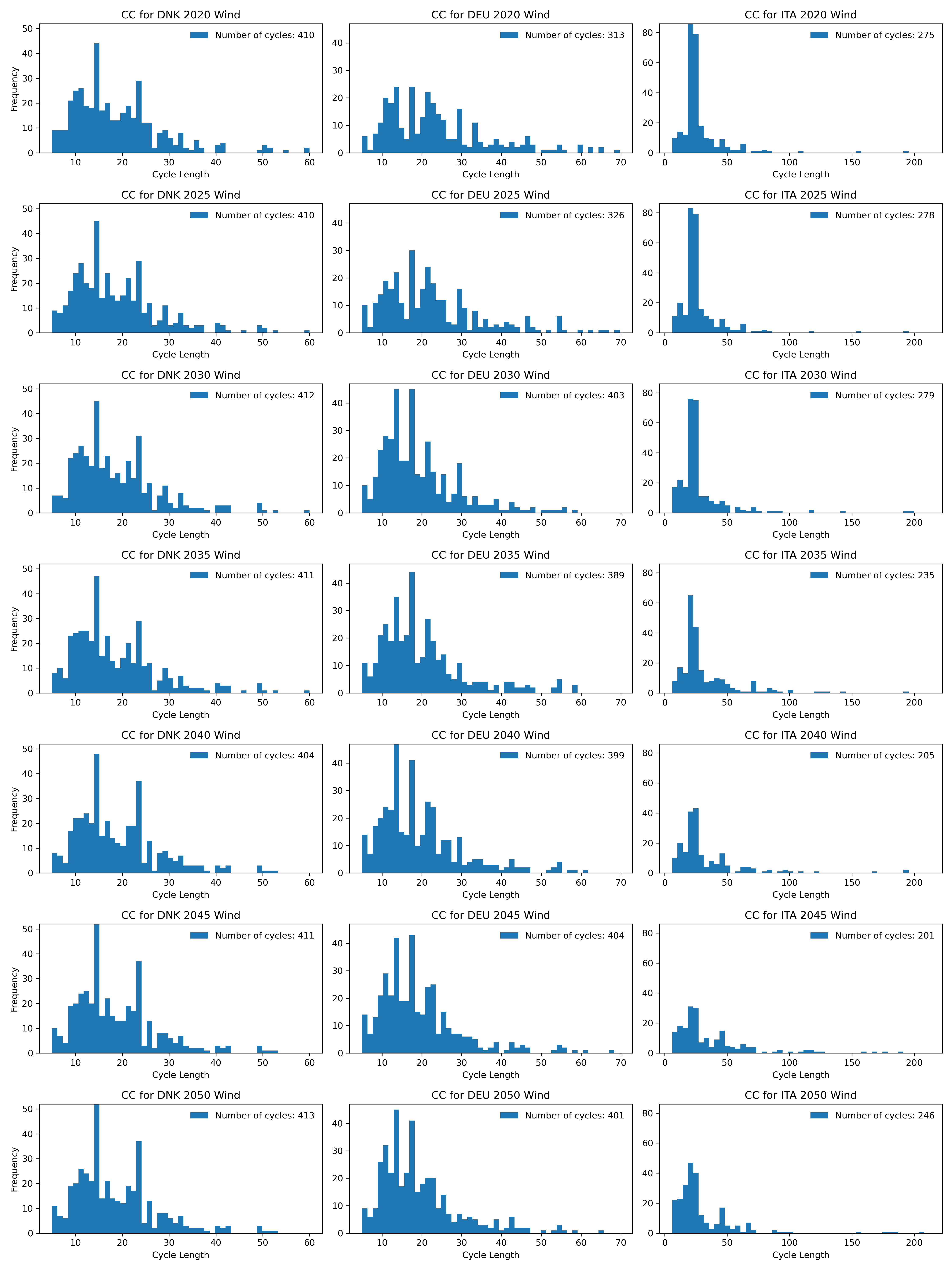}	
	\caption{Cycle Capture (CC) Analysis of Wind Generations in Denmark, Germany, and Italy} 
	\label{fig_S24}
\end{figure*}

\begin{figure*}[t]
	\centering 
	\includegraphics[width=0.9\textwidth, angle=0]{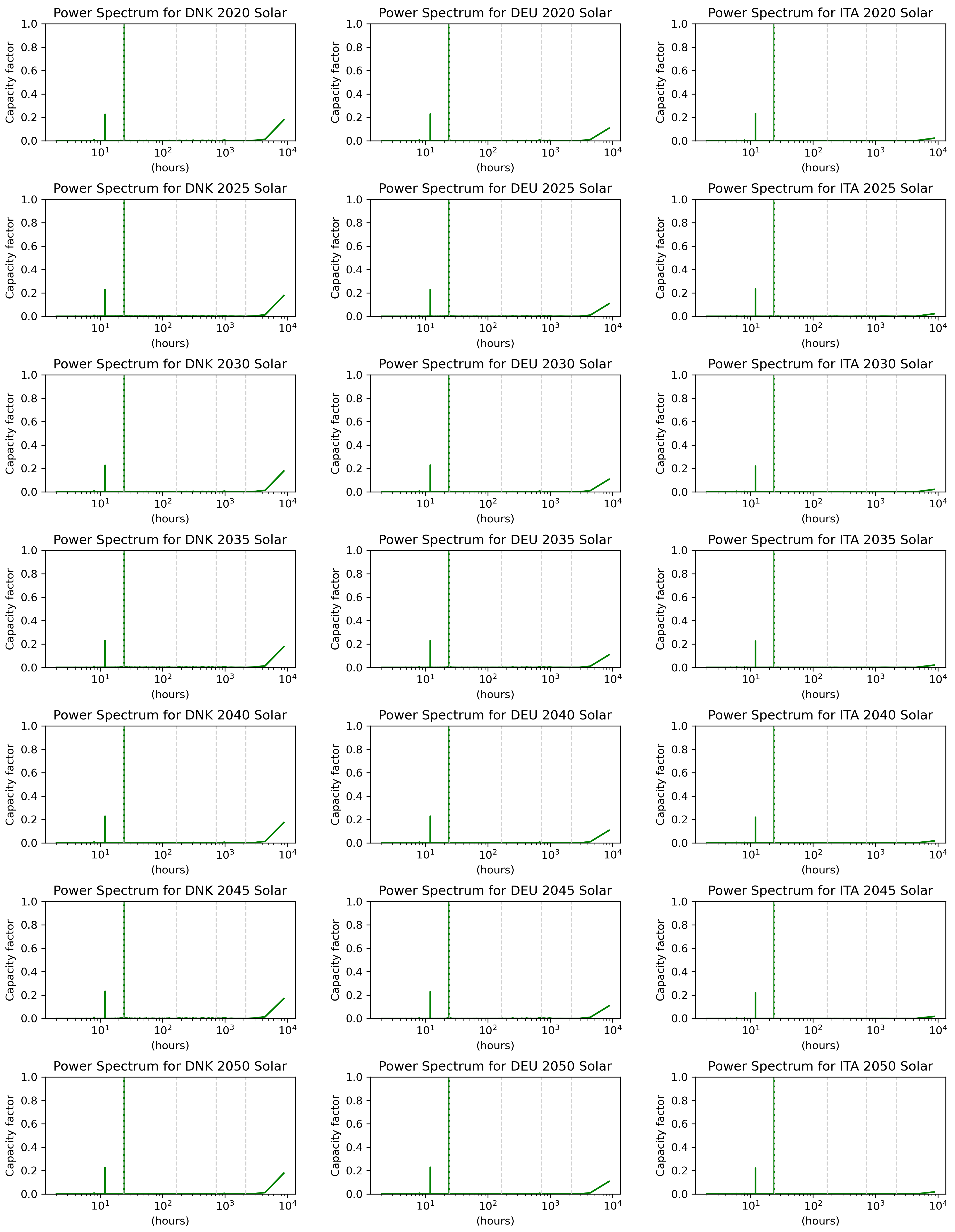}	
	\caption{Fast Fourier Transform (FFT) Analysis of Solar Generations in Denmark, Germany, and Italy} 
	\label{fig_S25}
\end{figure*}

\begin{figure*}[t]
	\centering 
	\includegraphics[width=.95\textwidth, angle=0]{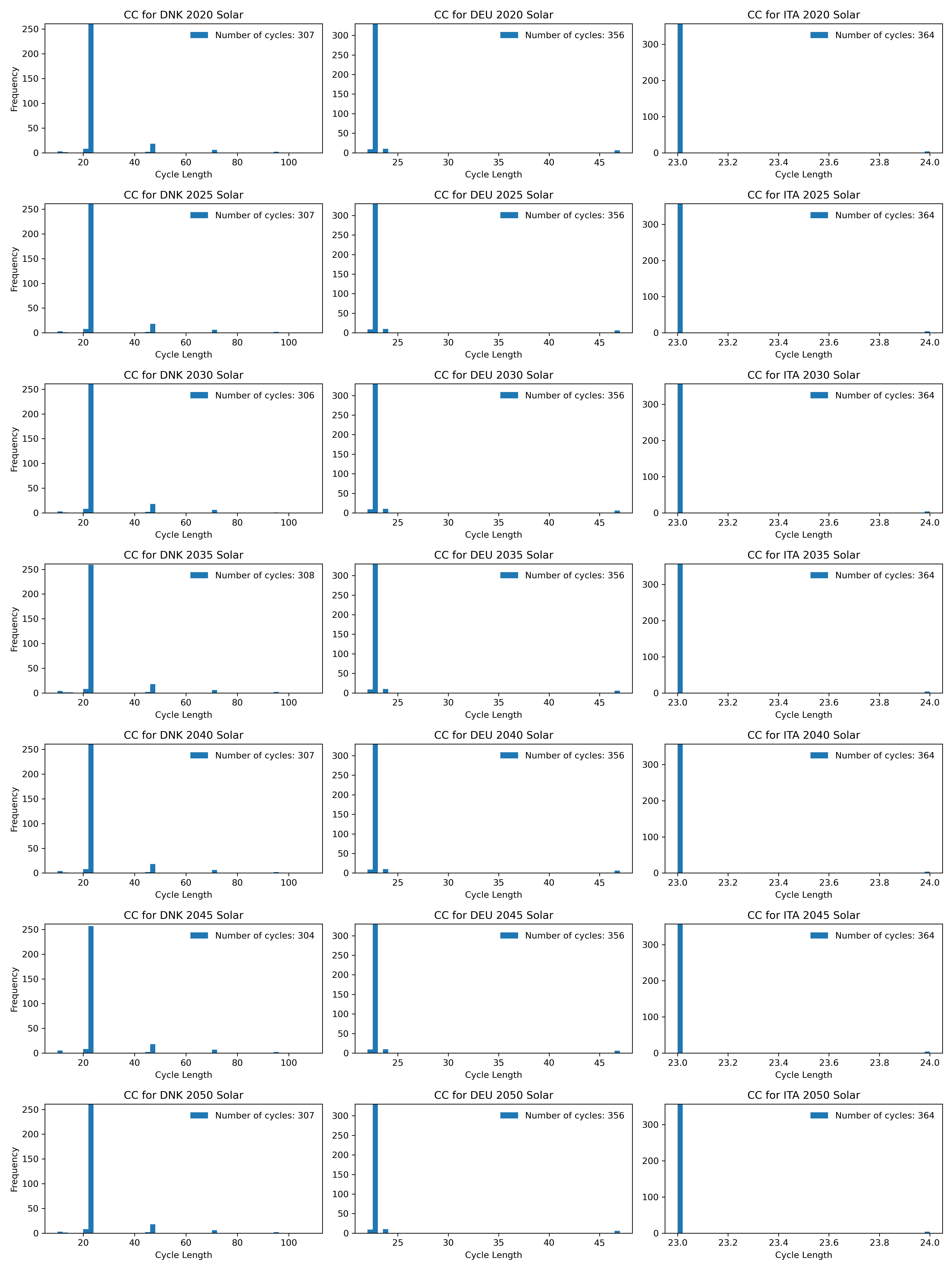}	
	\caption{Cycle Capture (CC) Analysis of Solar Generations in Denmark, Germany, and Italy} 
	\label{fig_S26}
\end{figure*}


\newpage
\bibliographystyle{elsarticle-num}
\bibliography{MainLib}






\end{document}